\documentclass{aa}  

\usepackage{graphicx}
\usepackage{lscape,color}
\usepackage{txfonts}
\usepackage{keyval}
\newcommand{\xmm}{{\it XMM}-Newton}

\newcommand{\swift}{{\it Swift}}

\newcommand{\nustar}{{\it NuSTAR}}
\newcommand{\hst}{{\it HST}}
\newcommand{\ms}{$M_{\odot}$}

\newcommand{\fluxcgs}{ergs~s$^{-1}$~cm$^{-2}$~\rm}
\newcommand{\lumcgs}{ergs~s$^{-1}$}

\newcommand{\tcrb}{T~CrB}
\newcommand{\nodata}{...}
\newcommand{\jls}[1]{{\color{cyan}{#1}}}

\begin{document}

\title{Evolution of the recent high-accretion state of the recurrent nova \tcrb: \hst, \swift, \nustar,\ and \xmm\ observations}

\titlerunning{Evolution of the recent high-accretion state of T CrB}

\authorrunning{G. J. M. Luna et al.}

\author{G. J. M. Luna
          \inst{1,2},
          N. P. M. Kuin \inst{3},
          K. Mukai \inst{4,5},        
          J. L. Sokoloski \inst{6},
          K. Page \inst{7}
          \and
          J. P. Osborne \inst{7}
          }

   \institute{Universidad Nacional de Hurlingham (UNAHUR). Laboratorio de Investigación y Desarrollo Experimental en Computación, Av. Gdor. Vergara 2222, Villa Tesei, Buenos Aires, Argentina\\
   \email{juan.luna@unahur.edu.ar}
   \and
   Consejo Nacional de Investigaciones Científicas y Técnicas (CONICET).
    \and
Mullard Space Science Laboratory, University College London, Holmbury St Mary, Dorking, Surrey RH5 6NT, UK.
\and
CRESST and X-ray Astrophysics Laboratory, NASA Goddard Space Flight Center, Greenbelt, MD 20771, USA.
\and
Department of Physics, University of Maryland, Baltimore County, 1000 Hilltop Circle, Baltimore, MD 21250, USA.
\and
Columbia University, Dept. of Astronomy, 550 West 120th Street, New York, NY 10027, USA.
\and
School of Physics and Astronomy, University of Leicester, Leicester LE1 7RH, UK}

   \date{Received 26 September 2025; accepted 18 January 2026}

\abstract
{As the recurrent nova T Coronae Borealis (\tcrb) approaches its next predicted thermonuclear eruption, it is currently exhibiting a "super-active state" (SAS) characterized by enhanced multiwavelength emission similar to the behavior recorded prior to the 1946 outburst. We present a multiwavelength analysis of the SAS and the subsequent "faint state" using observations from \hst, \swift, \nustar, and \xmm. Our results indicate that the SAS was driven by an increase in the mass accretion rate, which caused the accretion disk's boundary layer to become optically thick. A weighted least squares regression analysis quantifies the evolution of the accretion components, displaying a highly significant (4.5$\sigma$) increase in the luminosity of the optically thin cooling flow (L$_{cf}$) and a marginal (2.58$\sigma$) decrease in the optically thick boundary layer luminosity (L$_{bb}$) as the system transitioned into the faint state. We find that this dimming is consistent with an intrinsic change in the accretion flow rather than dust obscuration, supported by the lack of infrared excess and the stability of the 2175 \AA\ feature. Additionally, a time-series analysis using autoregressive modeling to account for correlated red noise revealed no significant periodicities, thereby disputing the previously reported $\sim$6000 s signal. These findings suggest that the pre-outburst evolution of \tcrb\ is characterized by significant changes in the accretion disk structure and boundary layer, providing a self-consistent physical framework for the system's behavior as it approaches eruption.}

\maketitle
%
%-------------------------------------------------------------------

\section{Introduction}

T~ Coronae Borealis (\tcrb) is a binary system that hosts a white dwarf and a M4III red giant (RG) on a 227.58 day orbit \citep{2025A&A...694A..85P}. The white dwarf (WD) has a mass of about 1.2 to 1.35 \ms, depending on the applied method \citep[see e.g.,][]{2008ASPC..401..342L,1998MNRAS.296...77B}. The accretion rate on the WD is variable within the range of below 10$^{-9}$ to above 10$^{-8}$ \ms\ yr$^{-1}$. When these parameters are combined, it creates the perfect recipe for describing recurrent nova outbursts. \tcrb\ is a recurrent nova with a roughly 80 year recurrence period, with eruptions recorded in 1866 and 1946 \citep[and also likely in 1217 and 1787,][]{2023arXiv230813668S} when it brightened so intensely as to reach the second magnitude. The next eruption is expected soon, with many authors proposing different dates (or range of dates), although its actual date is highly uncertain to predict \citep[see][and the discussion in Sect. \ref{sec:disc}]{2025MNRAS.541L..14M}.

The optical light curve obtained before and after the 1866 and 1946 nova outbursts showed a distinct brightness change ($\Delta V \approx$1, $\Delta B \approx$1.5) that started about 8 years before the nova outbursts and lasted about 7 years after the outburst, after being "interrupted" by a faint, one-year-long state (hereafter called the "Dip") and the nova outburst itself. A similar state started at the end of 2014 and was interrupted around October 2023 \citep{2023RNAAS...7..145M}. \citet{munaritcrb} coined the term ``super-active state'' (hereafter, SAS) to identify this state and we  use this term throughout the paper. Now that \tcrb\ has gone through the first portion of the SAS, the white dwarf may be close to having sufficient fresh fuel for a new nova eruption; however, the existing data do not allow a reliable prediction of its timing. During the SAS, the orbital modulation in the V-band light curve becomes less evident, the B-V color becomes bluer, and the flux of the emission lines increases \citep[see][]{Munari2025}. Shortly
after the onset of the 2014 SAS state, X-ray and ultraviolet (UV) observations  show the presence of a super-soft X-ray component, with a luminosity too low to be attributed to nuclear burning on the WD surface\footnote{We note that this is not a super-soft component in the sense of one associated with continued shell burning.}. \citet{tcrb2018} proposed that this component might have originated in the boundary layer of the accretion disk and, after an increase in the accretion rate, it became mostly optically thick to its own radiation; this scenario would be analogous to dwarf-nova, disk instability-type outbursts. At the same time, the X-ray emission in the 2--50 keV range ({\it hard} X-ray emission) faded, as a consequence of the much higher optical depth of the boundary layer caused by the increased accretion rate. In this scenario, once the disk returns to its quiescent state, it is expected that the boundary layer will evolve to a mostly optically thin regime. 

Most theoretical models of nova eruptions assume that mass should be constantly accreted at a rate of a few 10$^{-8}$ \ms\ yr$^{-1}$ on a massive WD, such as the one present in \tcrb,\  to support a nova eruption every $\sim$80 years \citep{2005ApJ...623..398Y}. \citet{2020ApJ...902L..14L} suggested that because most measurements of $\dot{M}$ outside the SAS yielded a few 10$^{-9}$ \ms\ yr$^{-1}$, then most of the mass required to trigger the nova eruption might be accreted during the SAS. The development of this state before the nova eruption might hold the key to our understanding of how novae accumulate mass during quiescence and what sets their recurrence time.

We describe the multiwavelength observations of \tcrb\ obtained with \hst, \swift, \nustar, and \xmm, along with studies of the historical optical light curve from AAVSO during the current development of the SAS through August 2025. In Section \ref{sec:obs}, we detail the steps we took to analyze the different datasets, while in Section \ref{sec:res}, we highlight our results. Section \ref{sec:disc} presents a discussion and our interpretations. The conclusions are presented in Section \ref{sec:conc}.

\section{Observations and data reduction} \label{sec:obs}

Figure \ref{fig:xrtlc} provides a summary of the variations in brightness observed in \tcrb\ since 2012 across various wavelengths. Prior to the onset of the SAS, which began in late 2014, \citet{tcrb2018} showed that \tcrb\ was typically detected by \swift-XRT at a rate of approximately 0.1 c s$^{-1}$. However, as Figure \ref{fig:xrtlc} shows, after the SAS commenced, the X-ray emission over the 0.3-10 keV range declined to only a few 10$^{-2}$ c s$^{-1}$ until early 2023, when it began to increase again as the dip phase initiated. A similar pattern has been observed at higher energies, such as those captured by \swift/BAT (14-50 keV), with a significant drop in flux during the SAS and a subsequent brightening upon its conclusion. In contrast, at lower energies (optical and UV), \tcrb\ experienced an increase in brightness during the SAS, a return to pre-SAS levels during the dip, and brightening again during the recovery phase, as illustrated in panels $(iii)$ and $(iv)$ of the figure. The light curve from the \swift/UVOT UV filters displays a similar, albeit more pronounced pattern compared to the optical data; notably, during the dip when \tcrb\ dimmed by over 2.5 magnitudes. The evolution of B-V, depicted in panel $(v)$, offers insight into the physics of these phases, which is explored in detail in the following sections.

\subsection{Optical photometry from AAVSO}

\begin{figure*}
\includegraphics[scale=0.35]{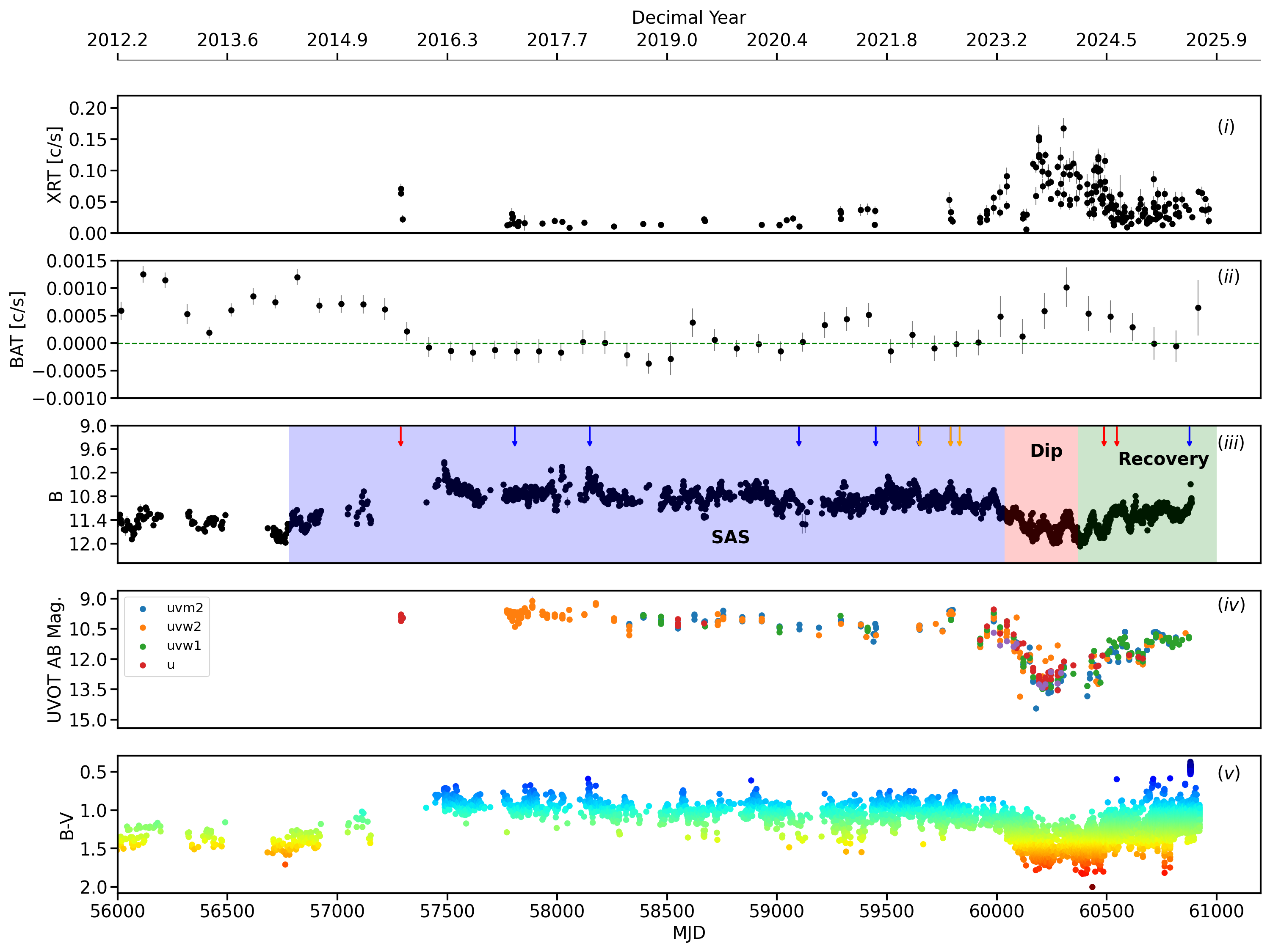}
\caption{Long-term, multiwavelength light curves of \tcrb\ since March, 2012 until August 2025. From top to bottom: ($i$)  \swift/XRT light curve in the 0.3-10 keV energy range; ($ii$) \swift/BAT 14–50 keV light curve with 100-day bins. 
($iii$) AAVSO B-band light curve (Johnson magnitudes). Downward arrows mark the dates of the \xmm\ (blue), \hst\ (orange), and \nustar\ (red)  observations. Note: Because some \xmm\ and \hst\ observations were quasi-simultaneous, their marks overlap in this panel and we only mark the dates of those \hst\ observations flagged as {\em Good} in Table \ref{tab:hst}. The purple-shadowed region comprises the period known as the SAS, while the pink-shaded region marks the period when the optical/UV brightness dropped to pre-SAS levels,  known as the dip, and the green-shadowed region shows the period that started around late 2023 when the brightness in optical/UV has started to rise again in the recovery phase; ($iv$) \swift/UVOT light curve in the $uvm2, uvw2, uvw1, u$ filters; ($v$) AAVSO B-V color, where red indicates a redder, cooler emission, while blue shows the opposite. }
\label{fig:xrtlc}
%\end{center}
\end{figure*}

We downloaded the B and V magnitude observations taken by the members of the AAVSO\footnote{from \url{https://www.aavso.org/data-download}}. In Fig. \ref{fig:xrtlc}, we present one-day average magnitudes and $B-V$ color since March 2012  to contextualize the optical flux state when the data from other wavelengths were taken.

\subsection{\hst}

T~CrB was observed with HST STIS to obtain medium-high resolution ultraviolet spectra at several epochs. Details of the observations are given in Table \ref{tab:hst}. The original goal was to obtain coeval HST/STIS observations with \xmm, but difficulties were encountered in executing some of the planned observations with the result that the target spectra were obtained over three periods (see Table~1). Of the failed spectra, the pointing was offset from the target, or the exposures were nulled out. 
The \texttt{\_x1d} files were spliced together using the IRAF\footnote{\url{https://iraf.net/}} \texttt{splice} tool.
Unfortunately, the flux calibration of some of the spectra is unreliable due to an offset of the target T~CrB; namely, those of 2022-07-30. However, this does not affect the line profiles with respect to the continuum.

%\renewcommand{\arraystretch}{1.3}
%\setstackgap{S}{0pt} 
\begin{table*}
\small
\caption{ \label{tab:hst} HST-STIS\ observations log. }
\centering
\begin{tabular}{lcccccr}
\hline\hline
STIS filter & Detector & ID  & Date & Exp. Time. & Orbital phase & Notes \\
\hline
E140M & FUV-MAMA & oeiv01010 & 2021-08-26 & 0      &0.68& failed\\
E230M & NUV-MAMA & oeiv01020 & 2021-08-26 & 0      &0.68& failed\\
E230M & NUV-MAMA & oeiv01030 & 2021-08-26 & 0      &0.68& failed\\
E140M & FUV-MAMA & oeiv02010 & 2022-03-11 & 1622.2 &0.56& good\\
E230M & NUV-MAMA & oeiv02020 & 2022-03-11 &  948.2 &0.56& good\\
E230M & NUV-MAMA & oeiv02030 & 2022-03-11 & 1284.2 &0.56& good\\
E140M & FUV-MAMA & oeiv51010 & 2022-07-30 & 1722.2 &0.18& good\\
E230M & NUV-MAMA & oeiv51020 & 2022-07-30 & 980.2  &0.18& off-target\\
E230M & NUV-MAMA & oeiv51030 & 2022-07-30 & 1316.7 &0.18& off-target\\
E230M & NUV-MAMA & oeiv52010 & 2022-09-09 &  600.2 &0.36& good\\
E230M & NUV-MAMA & oeiv52020 & 2022-09-09 &  600.2 &0.36& good\\
\hline
\end{tabular}
\end{table*}

\subsection{\swift}

\subsubsection{BAT}

The hard X-ray light curve in the 14--50 keV energy range was obtained from the \swift~ BAT Transient Monitor\footnote{\url{https://swift.gsfc.nasa.gov/results/transients/}}. We grouped the daily-binned light curve in 100-day-long bins to increase the signal-to-noise ratio (S/N) values in each bin. The BAT light curve is depicted in Figure \ref{fig:xrtlc}. %, where the ....

\subsubsection{XRT}

\swift\ obtained pointed observations with the X-ray Telescope (XRT) frequently during the last years through Target of Opportunity (ToO) requests by the authors, with approximately a monthly cadence, and most recently through a GO program (Prop. 2023261, PI. Sokoloski) on a roughly two-week cadence. Using the software package HEASoft v6.36, we extracted the source spectra and light curves from events with standard grades (0--12) within a circular region centered on \tcrb\ coordinates, while the background spectra and light curves were extracted from an annulus region centered on \tcrb\ coordinates with inner and outer radii of 40$^{\prime\prime}$ and 60$^{\prime\prime}$, respectively. The light curve presented in Fig. \ref{fig:xrtlc} shows the count rate per observation since March 2012.

\subsubsection{UVOT}

In addition to XRT observations, the Ultraviolet and Optical Telescope (UVOT) on \swift\ observed \tcrb\ using several modes: UV Grism, six-filter photometry, and timing in the $uvm2$ filter using the event mode. UV photometry observations started infrequently in 2004 with the first grism spectrum taken in October 2015; observations with the UV grism were regularly obtained since June 2018 with a 10d cadence. 
%and from 201x with frequent grism exposures. 
A reduction of the photometric data was done using aperture photometry when not too bright \citep[full frame count rates $<$ 0.97,][]{2008MNRAS.383..627P,2011AIPC.1358..373B}; however, when exceeding this limit up to brightness of 8.5 magnitudes,   the readout-streak photometry was used \citep{2013MNRAS.436.1684P}. The UV grism spectra were obtained using the UVOTPY programs \citep{2015MNRAS.449.2514K,2014ascl.soft10004K}.

\subsection{\xmm}

We analyzed seven observations obtained with \xmm\ (see Table \ref{tab:xmm_log} for the observing log) since 2017,  which are hereafter referred to as Obs1, Obs2, Obs3, Obs4, Obs5, Obs6, and Obs7. We used the XMM Science Analysis Software\footnote{\url{https://www.cosmos.esa.int/web/xmm-newton/sas}} (not to be confused with "SAS" used here) version 22.1.0. We extracted the spectra, light curves, and response matrices from the $pn$, MOS1, and MOS2 cameras as well as the RGS spectrometer from the source and background using the metatask \texttt{xmmextractor}. 
Overall, the spectra from the RGS do not have a high-enough S/N to allow for meaningful modeling; thus,  we did not include them in the spectral modeling. However, we did extract the RGS1+2 from Obs1 to search for the emission lines mentioned in \citet{2024MNRAS.532.1421T}. To this end, we used the script \texttt{rgsproc} to extract the spectra and \texttt{rgscombine} to combine the RGS1 and RGS2 gratings. 
The X-ray light curves from the EPIC-$pn$, MOS1, and MOS2 instruments were extracted and combined using a custom script that first determines the global time range of the observation by using \texttt{fkeyprint} to read the \texttt{TSTART} and \texttt{TSTOP} keywords from the headers of all three flare-filtered and barycenter-corrected event files; it then selects the absolute minimum start time and maximum stop times. For each instrument, the source and background light curves were then extracted from these event files using \texttt{evselect}, ensuring they were created on this common time grid.

The \texttt{epiclccorr} task was then applied to each instrument's light curves to perform the background subtraction and apply all necessary instrumental corrections (e.g., for vignetting, dead time, and quantum efficiency), producing three individual, corrected light curves. Finally, these three corrected light curves were co-added into a single, high-S/N EPIC light curve using the \texttt{lcmath} tool. The resulting FITS file contains the combined, background-subtracted count rate for the entire EPIC instrument suite using 100 s bins. The Optical Monitor (OM) observed both in fast and image modes with the UVM2 filter (except for Obs1, which used the UVW1, U, B, and V filters). In the case of fast mode, the light curves consist of the photometric series where the complete exposure time (see fourth column in Table \ref{tab:xmm_log}) was split into segments lasting approximately 4400 seconds and were extracted with a bin size of 10 seconds. The optical brightness during Obs1 led to high noise due to strong saturation effects in the OM fast mode data of the U, B, and V filters, while only two segments were useful in the UVW1 filter. During Obs7, the \xmm\ observing planners recommended blocking the OM to avoid possible damage to the OM photocathode in case of a sudden change in the V magnitude.

\begin{table*}
\small
\caption{ \label{tab:xmm_log} \xmm\ and \nustar\} observations log. }
\centering
\begin{tabular}{lcccccc}
\hline\hline
ObsID & Start Date & MJD & Exp. Time &  Mode/Filter & OM Mode/Filters& PI \\
\hline
0793183601 (Obs1) & 2017-02-23 UT03:47:34 &  57807.15 &63800  & Full/Medium & Fast/UVW1, U, B, V & DDT/G. Luna\\
0800420201 (Obs2)& 2018-01-30 UT04:25:43 & 58148.18 & 28000 & Full/Medium& Fast/UVM2 &Zhekov\\
0864030101 (Obs3)& 2020-09-07 UT23:50:02 & 59099.99 & 53800 & Full/Medium& Fast/UVM2 &J. Sokoloski\\
0882640301      (Obs4)& 2021-08-23 UT09:20:44 & 59449.38 & 64000 & Full/Medium& Fast/UVM2 &K. Mukai\\
0882640401      (Obs5)& 2022-03-08 UT09:52:59   & 59646.41 & 53000 & Full/Medium& Fast/UVM2 &K. Mukai\\
%0882640701     & 2022-07-29 UT13:58:44 & 13700 && Mukai\\
0882640601      (Obs6)& 2022-07-29 UT17:47:04   & 59789.74 & 56900      & Full/Medium& Fast/UVM2 &K. Mukai\\
0971190501 (Obs7)& 2025-07-21 UT02:51:50 & 60877.12 & 38000 & Full/Medium & bBocked & DDT/G. Luna \\
\hline
80601307002 (Nu1) & 2020-09-08 UT00:36:09 & 59100.02 & 49400 & \nodata & \nodata & J. Sokoloski \\
80901344002 (Nu2) & 2024-06-27 UT09:31:06 & 60488.39 & 66500 & \nodata & \nodata & K. Mukai \\
80901344004 (Nu3) & 2024-08-24 UT15:21:10 & 60546.63 & 31100 & \nodata & \nodata & K. Mukai \\
\hline
\end{tabular} 
\end{table*}

\subsection{\nustar}

We observed \tcrb\ with \nustar\ using both FPMA and FPMB modules on 2020-09-08 UT00:36:09 (Nu1), quasi-simultaneously with \xmm\ Obs3 and on 2024-06-27 09:31:06 (Nu2) and 2024-08-24 15:21:10 (Nu3) during the {\it XRISM} Performance Verification (PV) phase observation of \tcrb, which will be published in a forthcoming paper by the XRISM team. The spectra and light curves were extracted from event files selecting events from a circular regions of 70$^{\prime\prime}$ radii and 95$^{\prime\prime}$ radii for the source and background, respectively. The response and ancillary matrices were constructed using the script \texttt{nuproducts} under HEASoft v6.36\footnote{\url{https://heasarc.gsfc.nasa.gov/docs/software/lheasoft/}}.

\section{Data analysis, results, and interpretation} \label{sec:res}

In this section, we describe the analysis methods applied to the multiwavelength data from \xmm, \nustar, \swift, and \hst. We then present the primary results and their interpretation from this analysis, which form the basis for the discussion in Section 4.

\subsection{Analysis methods}

\subsubsection{X-ray spectral fits} \label{sec:res_spec}

In all but one \xmm\ observation (Obs7), the X-ray spectrum can be modeled with three emitting components. At energies above $\sim$0.7 keV, the highly absorbed X-rays with the presence of emission lines in the 6--7 keV range arise from a cooling flow \citep[see e.g.,][]{2003ApJ...586L..77M}. Below 0.7 keV, the spectrum is dominated by a soft, blackbody-like component. The emission line at 6.4 keV, Fe K$\alpha$, originates in the reflection of the X-ray emission either in the white dwarf or accretion disk \citep{2019ApJ...880...94L}. Following \citet{tcrb2018}, we modeled the spectra with a model consisting of full and partial covering absorbers that modify a blackbody, a thermal component of cooling flow, and an emission line at 6.4 keV. While the temperature of the cooling flow plasma traces the gravitational potential well of the white dwarf, its normalization provides the mass accretion rate scaled by the distance, $\dot{M}_{thin}$, of the optically thin portion of the accretion disk. The luminosity of the blackbody-like component in turn can be used to estimate the mass accretion rate of the optically thick component of the boundary layer, $\dot{M}_{thick}$; thus, we have $\dot{M}$=$\dot{M}_{thin}$ + $\dot{M}_{thick}$ \citep[see][]{tcrb2018}. It is crucial to clarify the physical interpretation of the parameters derived from our spectral fits. We modeled two distinct, luminous regions that both respond to the overall mass transfer rate onto the white dwarf ($\dot{M}$). The luminosity of the optically thick component (L$_{BB}$) traces the accretion energy released in the dense, optically thick portion of the boundary layer. In contrast, the luminosity derived from the cooling-flow model (L$_{CF}$) traces the energy released in the hot, optically thin portion of the same boundary layer. These are not independent or competing accretion rates; rather, they are distinct luminous tracers of different parts of the same complex accretion flow. By tracking both, we can diagnose how the structure of the accretion region evolves as the overall mass transfer rate changes. To quantitatively assess the trend in the luminosity (L$_{BB}$) of the optically thick and thin components throughout the SAS, dip and recovery phases, we performed a weighted least-squares (WLS) regression of L$_{BB}$ and L$_{CF}$ versus MJD, using the inverse variance of the measurement errors (1/$\sigma^2$) as weights (see Fig. \ref{fig:xmm_evol}). 

The \nustar\ 2024 spectra (Nu2 and Nu3) were modeled together with \swift\ observations obtained quasi-simultaneously (\swift\ ObsIDs 00097564043, 00097564044, 00097564045, 00097564046, 00097564047, and 000975640438, together with Nu2, and additionally ObsIDs 00089722001 with Nu3). In the \nustar\ observations (Nu2 and Nu3) and the last \xmm\ observation (Obs7), all taken after the interruption of the SAS, the presence of a blackbody component cannot be constrained; thus, the model we fit consists of full and partial covering absorbers that modify a cooling flow thermal component and an emission line at 6.4 keV (Fe K$\alpha$). We note that the fit for Obs7 shows some minor, unmodeled residuals around 2 keV (see Fig. \ref{fig:all_spec}). These might indicate the presence of weak, complex emission features that were not captured by our simple model. Nonetheless, they are not statistically significant and do not affect our overall conclusions regarding the non-detection of the blackbody component.

\subsubsection{Timing analysis} \label{sec:timing}

According to \citet{zhekov19}, a modulation with a period of around 6000~s was identified during the Obs1 and Obs2 sessions. This periodicity was established using the Lomb-Scargle (LS) periodogram analysis of the soft X-ray band (0.2--0.6 keV). This soft component was detected at a similar intensity during Obs5, while it was significantly weaker during the Obs3, Obs4, and Obs6. 

The light curves in the 0.2-0.6 keV energy range and the OM data show strong variability, resembling what is normally referred to as red noise; thus, the assessment of any significance level for period detection must take this into account. To search for periodic signals, we performed a time-series analysis designed to account for the presence of correlated (red) noise, which is common in accretion-driven systems. The presence of red noise can artificially inflate the significance of low-frequency peaks in a standard periodogram, leading to false positives. Our analysis proceeds in three main steps, implemented using the Python packages \texttt{Astropy} \citep{astropy:2013, astropy:2018} and \texttt{statsmodels} \citep{seabold:2010}. 

First, we characterized the nature of the correlated noise in the mean-subtracted light curve. We computed the autocorrelation function (ACF) and partial autocorrelation function (PACF) of the time series. The slow, exponential decay observed in the ACF is characteristic of an autoregressive (AR) process, where the flux at a given time is correlated with the flux at previous times.

Second, to model and remove this red noise, we fitted an autoregressive model of the order of $p$, AR($p$), to the light curve. The optimal order $p$ was determined by fitting models with increasing complexity (from $p$=1 to $p$=10) and selecting the order that minimized the Akaike information criterion (AIC). This data-driven approach ensures that the model is complex enough to capture the underlying red noise structure without overfitting the data. The best-fit AR($p$) model represents the stochastic, correlated variability in the light curve. 

Finally, we subtracted the best-fit AR($p$) model from the data to produce a set of residuals. The ACF and PACF of these residuals were examined to ensure they are consistent with uncorrelated (white) noise, confirming that the model successfully accounted for the red noise component. We then computed a LS periodogram \citep{vanderplas:2018} on these "whitened" residuals to search for any significant, coherent periodicities that were not attributable to the stochastic red noise. The significance of any peaks in the residual periodogram was evaluated by calculating the false alarm probability (FAP) at the 99.99\% confidence level, which gives the probability that a peak of a given power could arise from pure white noise.

\subsubsection{\hst\ spectroscopìc analysis \label{sec:timing}}

The HST spectra were downloaded from the MAST, the latest reprocessing used was on 2022-09-28. The interpretation of the spectrum utilized the NIST Atomic Spectroscopy Data\footnote{\url{https://www.nist.gov/pml/atomic-spectra-database}}.

\subsection{Results and interpretation}

\begin{figure*}
\begin{center}
\includegraphics[scale=0.6]{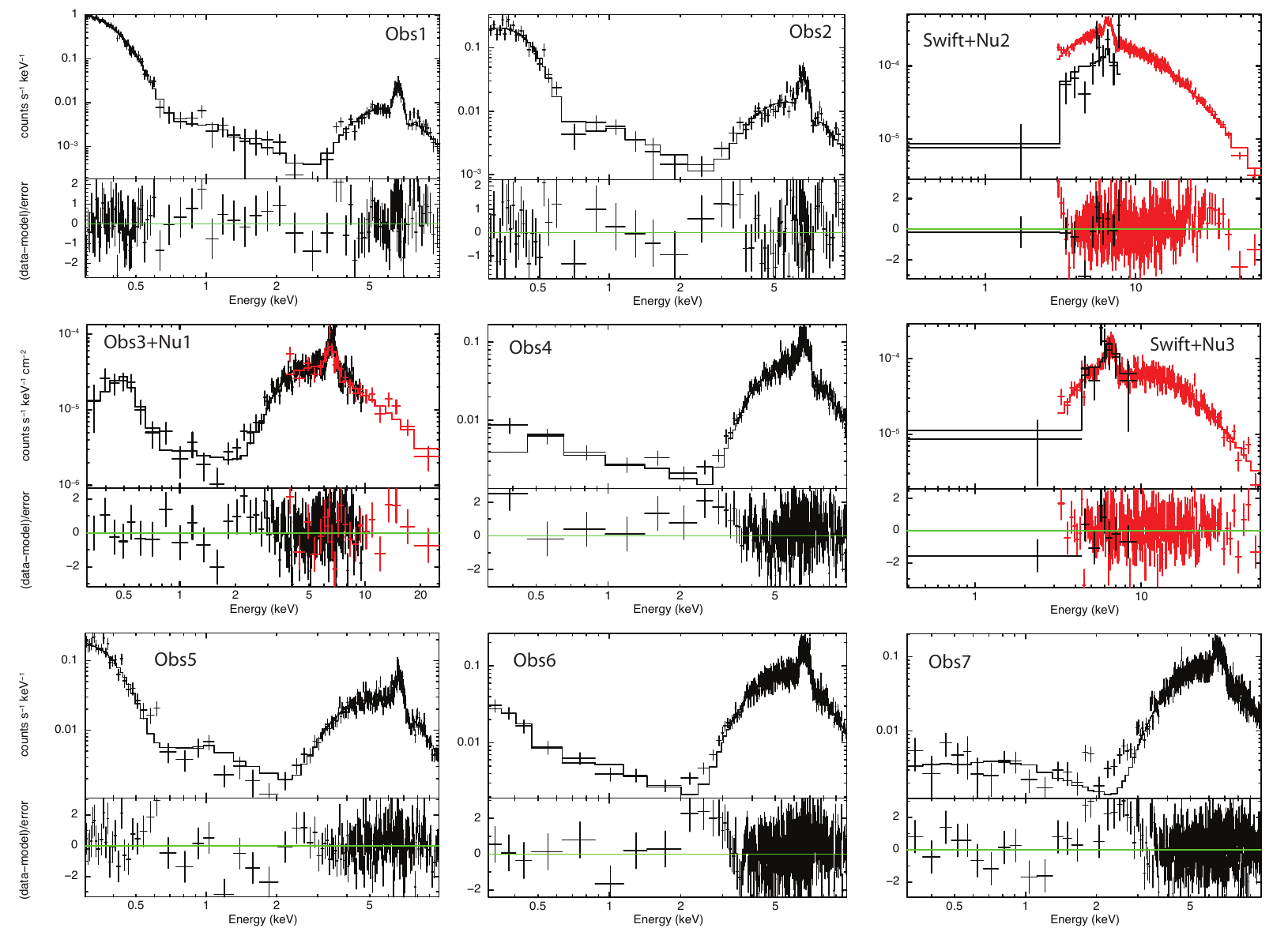}
\caption{{\it First and second columns:} \xmm/pn and \nustar\ (in red) spectra of T~CrB taken during the SAS. The model parameters, listed in Table \ref{tab:1}, were obtained by simultaneously fitting the pn, MOS 1 and MOS 2 spectra during each pointing. Here we show only the $pn$ spectra for clarity. {\it Third column:} \nustar/FPMA, \xmm,\ and \swift\ spectra of T~CrB taken during the faint state after the SAS. The \swift\ spectra were grouped at one count per bin and modeled using Cash statistic. 
}
\label{fig:all_spec}
\end{center}
\end{figure*}

\begin{figure*}
\begin{center}
\includegraphics[width=\linewidth]{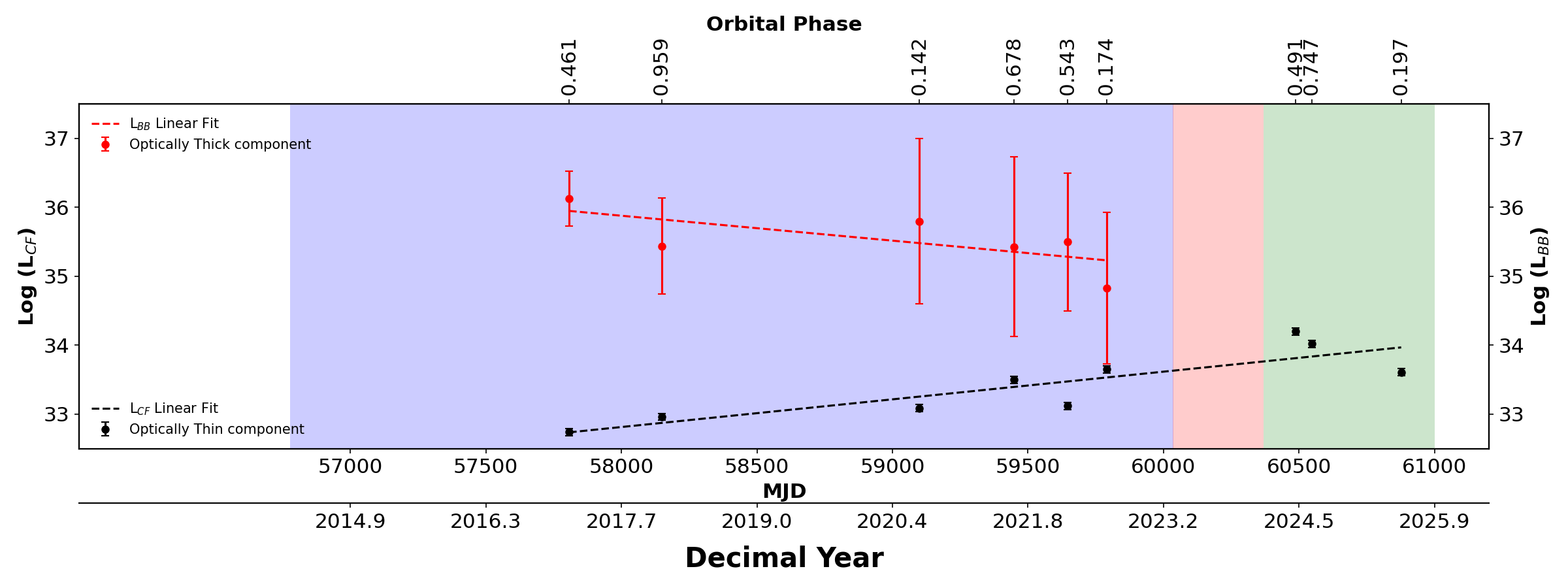}
\caption{Evolution of the luminosity from the black-body (red points) and hard X-ray emitting, cooling flow (black points) spectral components over time. The shaded areas mark the SAS, Dip and Recovery phase using the same color code as in Fig. \ref{fig:xrtlc}. A WLS fit to the luminosity of the optically thick (BB) component (dotted red line) indicates that its slope is different from zero at the 2.58$\sigma$ level. We include Nu2 and Nu3 observations taken in 2024 and Obs7 taken in 2025, whose spectra arise from the optically thin portion of the boundary layer, i.e., the cooling flow. Dotted black lines mark the WLS fit to the $L_{CF}$ evolution whose slope is different than zero at the 4.49$\sigma$ level. The increase in its luminosity suggests that the accretion rate in the disk decreased over time, with the boundary layer becoming optically thinner and hotter. }
\label{fig:xmm_evol}
\end{center}
\end{figure*}

Figure \ref{fig:all_spec} shows the \xmm-$pn$+\nustar\footnote{During the fit of each \xmm\ observation, we simultaneously modeled the $pn$ and MOS cameras. For clarity, Fig. \ref{fig:all_spec} includes only the $pn$ spectrum.} spectra together with the best-fit models and residuals for those spectra taken during the SAS. Figure \ref{fig:all_spec}, in turn, shows the spectra with best-fit models and residuals from the observations taken during the period of recovery (see green-shaded area in panel $iii$ of Fig. \ref{fig:xrtlc} for reference). The evolution of the luminosity of the optically thick (L$_{BB}$) and optically thin components (L$_{CF}$; $E>0.7$KeV) is depicted in Figure \ref{fig:xmm_evol}. The top X-axis also shows the corresponding orbital phase. First, we observe that any change in the luminosity of both components is not related to the binary orbital motion. Moreover, we notice a distinct trend of increasing luminosity in the optically thin component during the SAS, with even higher luminosity observed during the two \nustar\ observations conducted in the recovery phase.
During the most recent observation with \xmm, Obs7, taken also during the recovery phase, the luminosity of the cooling flow was seen to decrease. If this optical brightening, hard X-ray fading phase is akin to the rise toward the SAS witnessed in 2015 \citep{2019ApJ...880...94L}, it is possible that a new, super-soft luminous optical thick component will be detected within the next few months. 

The analysis of luminosity of the optically thick component (its area and temperature) reveals a decreasing trend with time with a slope that is 2.58$\sigma$ different from zero ($p$-value $\approx$ 0.01), as shown in Fig. \ref{fig:all_spec}. While the uncertainties on individual measurements are considerable, this result indicates that the overall decrease in luminosity over the $\sim$6 years of observation is statistically significant at a $\sim$99\% confidence level. We therefore conclude that the blackbody component, while remaining luminous, did experience a modest decline until the interruption of the SAS around April 2023. Furthermore, our analysis of the evolution of the optically thin, cooling flow component (L$_{CF}$) reveals a strong, statistically significant brightening trend throughout the SAS. The WLS fit shows the slope of the trend is positive and 4.49$\sigma$ away from zero ($p$-value $\approx$ 0.003). The seemingly opposite evolution between the luminosities of the super-soft and hard X-ray components reflects a change in optical depth of the X-ray emitting plasma in the boundary layer, directly related to changes in $\dot{M}$ (see Sect.~\ref{sec:disc}).

\begin{figure}
\begin{center}
\includegraphics[scale=0.4]{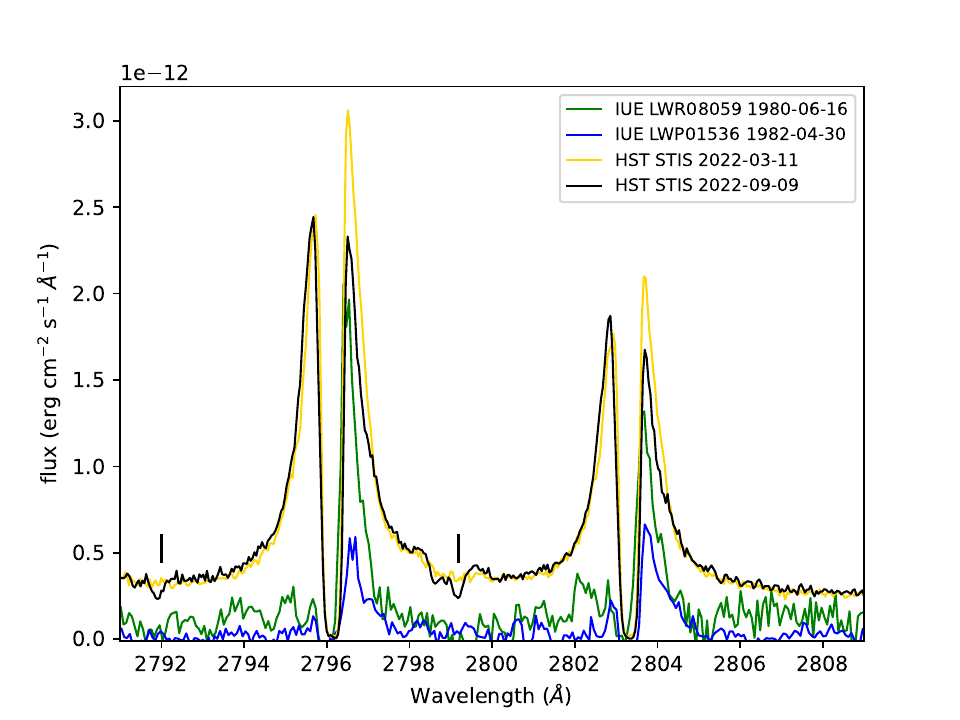}
\caption{Mg II doublet.  IUE spectra from before the SAS are plotted together with the HST STIS spectra from the end of the SAS period. Two small black lines are added to guide the eye to the absorptions that match the two Mg II profiles:\ blueshifted outflow in the line of sight for the 2022-09-09 (MJD59469, phase 0.56) spectrum, but not in the 2022-09-09 (MJD 59831, phase 0.36) spectrum. Note: the IUE Mg II spectra seem to be missing the blue wing. The background during SAS is higher than in the IUE spectra. The flux scale is listed above the left corner of the image.  }
\label{fig:mgII}
\end{center}
\end{figure}

\begin{figure*}
\begin{center}
\includegraphics[scale=0.7]{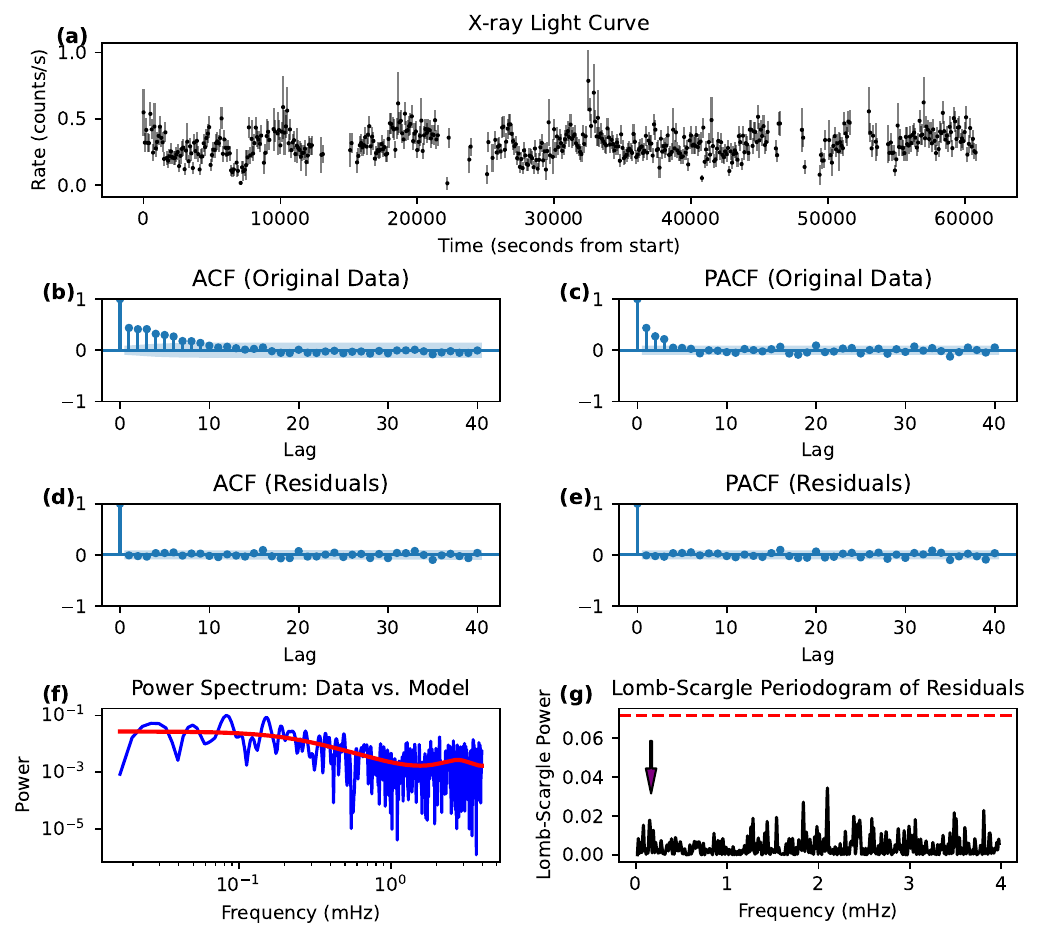}
\caption{Time-series analysis of the Obs1 \xmm/$pn$+MOS1+MOS2 light curve of \tcrb. (a)  Full, original light curve, showing significant aperiodic variability (red noise) across the observation. (b)  ACF of the original mean-subtracted data. The slow decay confirms the presence of strong, correlated red noise. (c) PACF of the original data. The significant power concentrated in the first few lags suggests that an autoregressive (AR) process is a suitable model for the red noise. (d)  ACF of the residuals after subtracting the best-fit AR(p) model. The lack of significant correlation spikes indicates that the model has successfully "whitened" the data, removing the dominant red noise component. (e)  PACF of the residuals. Similarly to the residual ACF, the absence of significant spikes confirms the goodness-of-fit of the AR(p) model. (f) log-log plot of the LS power spectrum of the original data (blue), which shows a classic red-noise profile with power increasing toward lower frequencies.  The theoretical power spectrum of the fitted AR(p) model (red) are overplotted, demonstrating an excellent match to the stochastic properties of the light curve. (g)  LS power spectrum of the "whitened" residuals. This plot should reveal periodic signals that were previously buried under the red noise. The horizontal dashed line marks the 99.99\% FAP level. The most prominent peak does not rise significantly above this threshold, indicating that its periodicity is not statistically significant. The arrow marks the frequency corresponding to the 6000 s period reported by \citet{zhekov19}.}
\label{fig:pn_2017}
\end{center}
\end{figure*}

The \hst\ STIS spectra show a multitude of emission lines formed with likely contributions both of the wind of the red giant (RG) and of the accretion disk. Here, only the Mg\,II lines at 2796.34 and 2803.52 are shown in Fig.~\ref{fig:mgII} with the IUE high resolution spectra from 1980 and 1982 long before the SAS. The central absorption is due to the interstellar medium. The difference in orbital phase leads to differences in the red or blue wing strength due to differences in outflows that scatter the photons. We also note the marked absorption dips on the blue wings: one for each doublet line at a velocity of -445~km/s in the 2022-09-09 phase 0.36 spectrum. This velocity is probably due to a stream intersecting the line of sight and causing the extinction and not from the wind. In these spectra, taken well within the SAS, the wind velocity from the width of the Mg II lines is around 100-140 km $s^{-1}$. Comparing with historical IUE spectra, the line width of Mg II is larger. The IUE spectra were taken from the INES archive \citep{2000Ap&SS.273..155W}, namely, high-resolution spectra LWR08059, LWP01536, and SWP11087. In the IUE SWP11087 (1981-01-17 ) high-resolution spectrum, we can determine the full width at half maximum (FWHM) line widths of several lines: C~I 1274.1, O~II 1317.1, Fe~II 1672.3, and 1810.3, 1825.3 $\AA$ with associated velocities of about 25 km $s^{-1}$. In the STIS spectrum, these lines are not observed; instead, we turn to the emission lines from higher excitation states. The FWHM of the resonance line of N~IV 1486 \AA~ indicates a velocity of 97~km $s^{-1}$; many other lines are broader, those lines suggest larger velocities but blends make estimates unreliable. The SAS UV spectrum is just very different, showing higher ionization lines than in the IUE epochs. The different line widths, despite the difference in ionization,  reveal that during the SAS, the wind of the secondary (and likely that of the accretion disk) has become quite different than that during quiescence, supporting the notion of an increased mass transfer rate. The correlated changes in UV line and continuum flux (see Fig. \ref{fig:mgII_trail}) suggest these arise from a physically connected source, such as an atmosphere with both optically thick and thin components.

A comprehensive search for coherent periodicities was conducted on both the \xmm/$pn$+MOS1+MOS2 and Optical Monitor light curves. Our analysis reveals no statistically significant periodicities in either the X-ray or optical data within the frequency range searched. Specifically, we found no evidence of the previously reported ~6000-second periodicity (see Fig. \ref{fig:pn_2017} and Appendix \ref{appen} figures). This null result suggests that during this observation, the variability in \tcrb\ was dominated by stochastic accretion processes, rather than by a stable, coherent clock.

\section{Discussion} \label{sec:disc}

\subsection{The faint state after the SAS.}

\begin{figure}
\begin{center}
\includegraphics[width=\linewidth]{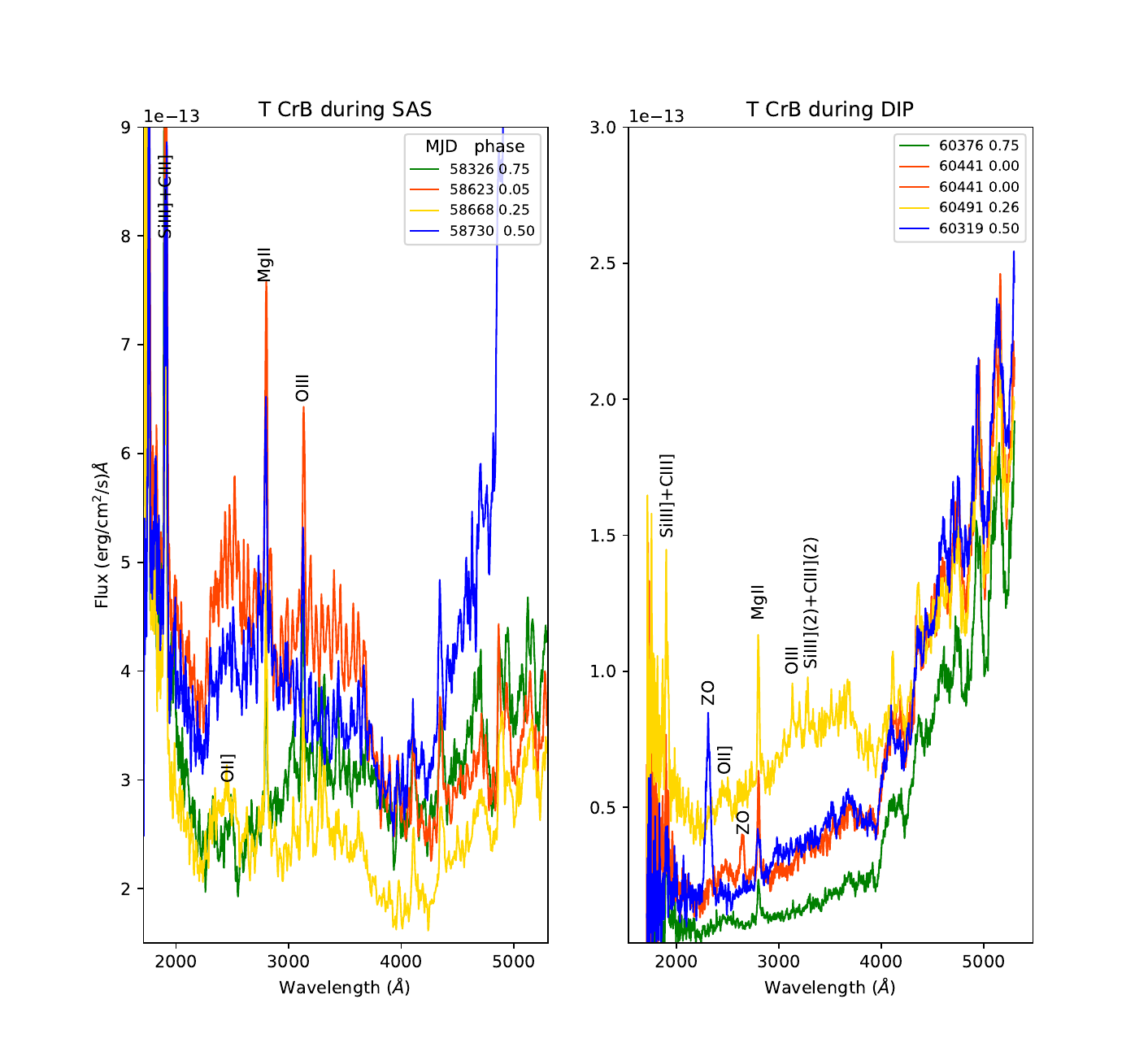}
\caption{ \swift\ UV spectra taken close to orbital phases 0.75, 0.0, 0.25, and 0.5 (velocity phase; WD occultation at 0.75). The left panel shows that during the SAS, the Balmer jump went into emission, likely due to scattering in the accretion disk. However, the dust absorption dip around 2150$\AA$ is not more pronounced after SAS ends.  During the dip after the SAS, the overall flux was much lower and there was no significant additional emission in the 2300-3600$\AA$ range. The blue dip (phase 0.5) spectrum shows contamination from a zeroth order around 2310$\AA$\ and in the orange (phase 0) spectrum around 2650$\AA,$\ which should be ignored. In the SAS spectra, we see second-order lines of (N~III](2) at 2900$\AA$, the Si~III](2), and C~III](2) lines at 3290 and 3320$\AA$). Fe~II UV triplet absorptions are present at 2260, 2600$\AA$. Finally, H~I 4105 and 4360$\AA$, as well as Mg~II 2800, C~III]~1909, and Si~III]~1892 $\AA$~ are in emission. The flux scale is at the top-left of each figure. }
\label{fig:uv_spec}
\end{center}
\end{figure}

\begin{figure}
\begin{center}
\includegraphics[width=\linewidth]{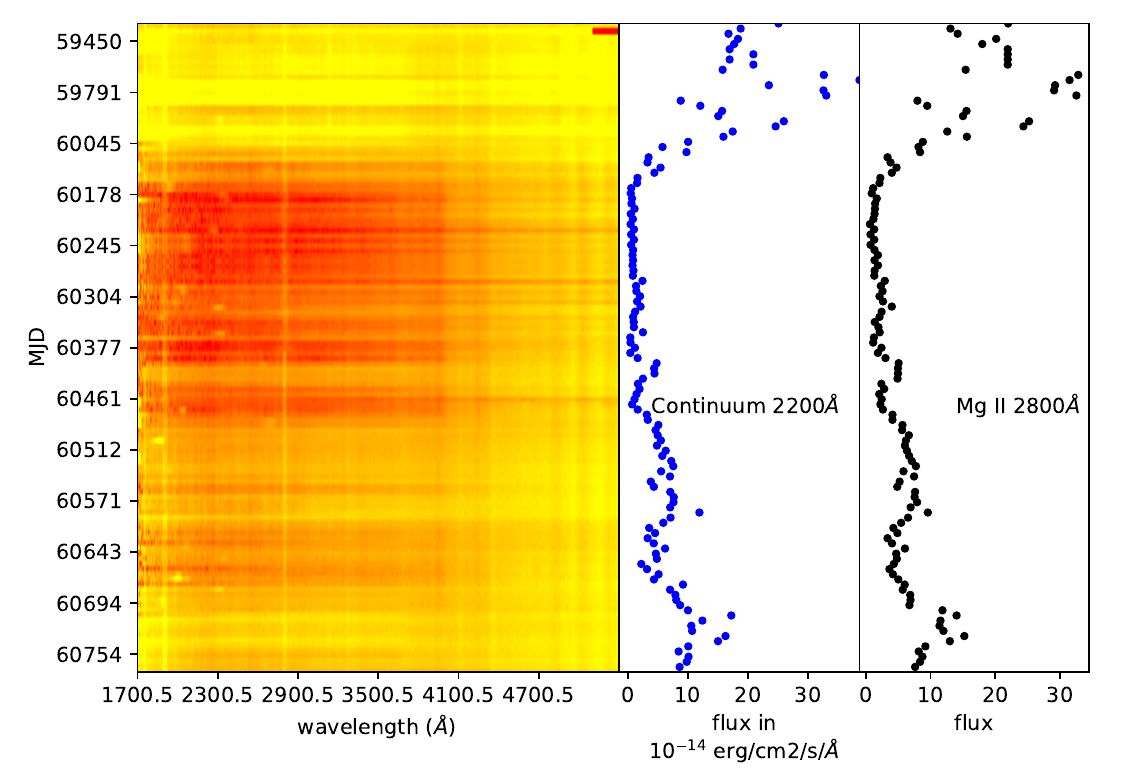}
\caption{Evolution of UV emission during and after the dip. The left-hand panel is an image composed of 127 \swift\ grism UVOT spectra taken from the end of the SAS period through the dip and the recent uptick in brightness. The intensity varies from yellow to red with yellow the brightest, red the weakest. With the exception of the pre-dip times and \swift\ observing restrictions, the spectra were typically taken around every 10 days. Some bright spots in the spectra are due to contaminating zeroth orders from field sources. Besides the dip, bright emission lines, the blend of Si~III]/C~III] (1892 and 1909 \AA), and Mg~II (2800 \AA) are seen as vertical yellow lines, on the red side the spectrum dominated by the red giant with its HI and molecular lines. This panel should also be compared with Fig.~\ref{fig:uv_spec} for context. In addition, the panel in the middle shows the continuum variation at 2200$\AA$ and highlights its behavior during the dip. In all three panels, the time axis is the Y-axis. The rightmost panel shows the net flux in Mg~II 2800$\AA$ which is formed partially in the low chromosphere and/or wind of the RG and partially in the accretion disk. The flux in the emission lines, e.g., Mg~II,  closely follows the UV continuum variations; this was discovered with IUE by \citet{1994ApJS...94..183M}. } 
\label{fig:mgII_trail}
\end{center}
\end{figure}

Before the previous two outbursts and since October 2023 \citep{munari23}, \tcrb\ entered a faint state, with V reaching pre-SAS levels. This state has been referred to in the literature as pre-eruption dip (and it is shown in Fig. \ref{fig:xrtlc}, marked as "dip")  because a similar state seems to have preceded the 1866 and 1946 eruptions and might perhaps lead to the next nova outburst \citep{2023arXiv230804104Z, tcrb_prediction, wagner_2025}. \cite{2023MNRAS.524.3146S} proposed that this state could be due to dust extinction from a previous ejection from the accretion disk. Two independent pieces of evidence suggest this hypothesis to be most unlikely: $i$) the lack of infrared (IR) emission from newly formed dust as reported by \cite{2023ATel16120....1W} and $ii$) the strength of the dust absorption-sensitive feature at 2150 \AA.  In Figure \ref{fig:uv_spec}, we compare the UVOT spectra taken on different orbital phases well into the SAS, and those spectra taken well within the so-called dip. In the near-UV (NUV), over 2000-3500 \AA, the spectrum contains many lines from singly ionized species, such as Fe\,II, and can show enhanced emission from the disk \citep[see, e.g.,][]{skopal2005A&A...440..995S} as well as absorption from dust (if present) around 2150 \AA. It can be seen that the feature at 2150 \AA,\ whose strength is proportional to the amount of absorbing material, was deeper with respect to the nearby emission during SAS, contrary to the expectation from the scenario of the dip as due to increased absorption. Moreover, looking at the flux variation over the orbit both during and after the SAS, there is evidence of some change, likely due to the extra emission during SAS; essentially, there is continued variability seen in the faint state after the SAS (see also Fig.~\ref{fig:mgII_trail}). If dust formation caused the dip state, we would expect the amplitude of the UV emission to decrease because the dust would then have to be system-wide. The B-V light curve in the bottom panel of Fig.~\ref{fig:xrtlc} shows that after SAS, \tcrb\ is as red as it was during the quiescent phases; this is contrary to the expectation for when dust is present, as it is re-emitting in the IR and causing a change in color toward the red side.

Alternatively, \citet{wagner_2025} proposed that the faint state before the 1866 and 1946 eruptions and after the SAS is due to the increase of the inner disk radius during a convective phase leading to the thermonuclear runaway.  \citet{teyssier} in turn, concluded that the pre-eruption dip is just the end of the SAS (i.e., the accretion rate returning to quiescent levels).
Given the recent claim by \citet{2025MNRAS.541L..14M} that the predicted thermonuclear outburst might be delayed, the causal connection between the faint state after the SAS and a subsequent thermonuclear runaway is being challenged.

\subsection{The origin of the softest X-rays}

\begin{figure}
\includegraphics[scale=0.4]{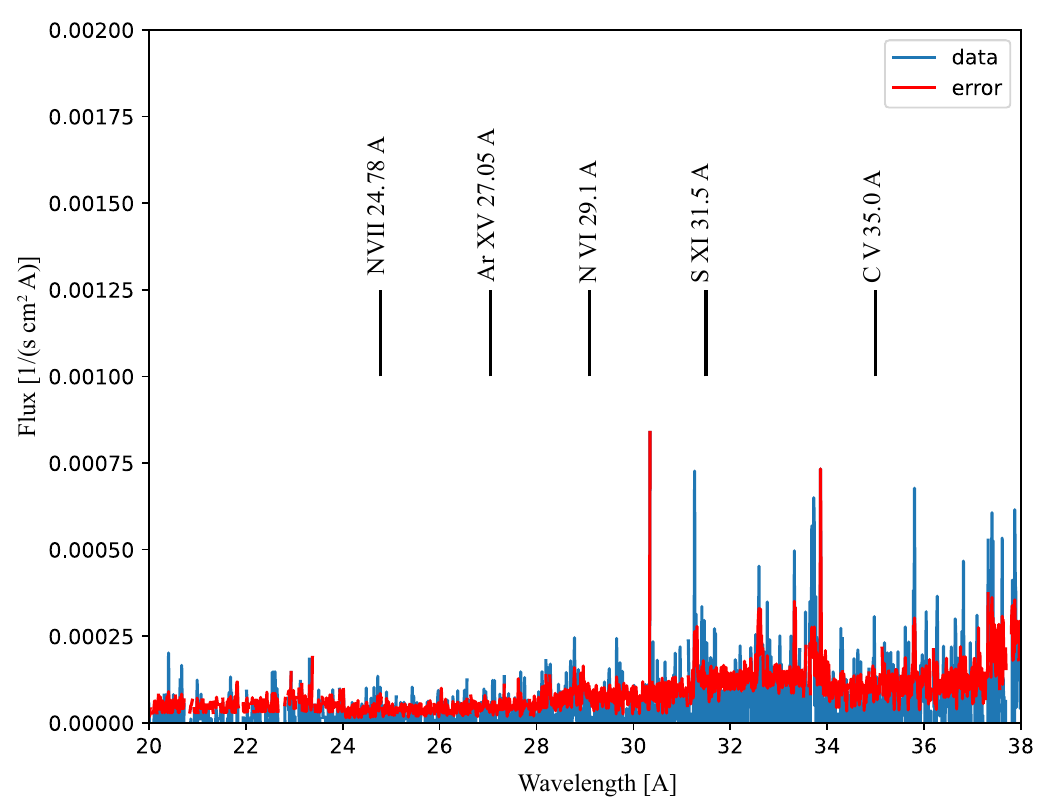}
\caption{\xmm/RGS1+2 spectrum (blue) and measurements errors (red) from the 2017 observations (Obs1). We show the position of the emission lines identified by \citet{2024MNRAS.532.1421T}, which can not be identified in the data over the background.}
\label{fig:rgs}
\end{figure}

\citet{2024MNRAS.532.1421T} proposed an alternative interpretation of the X-ray spectra obtained during the SAS with \xmm\ in which the softest X-rays arise from an optically thin thermal plasma (\texttt{apec}). This assertion was based on the RGS data of the \xmm\ 2017 observation (Obs1), where the authors identified emission lines that are not typically expected in the case of black body, optically thick emission. Overall, our analysis of those data indicates that the detection of emission lines above the background cannot be confirmed. Figure \ref{fig:rgs} shows the RGS1+2 spectrum and measurement errors together with the lines identified by \citet{2024MNRAS.532.1421T}. Lines cannot be identified in the RGS spectrum because of the faint signal that is overwhelmed by the background.

Furthermore, the absorption column of $n_H$=(2.3$\pm$0.1)$\times$10$^{21}$cm$^{-2}$ derived from their fit of the EPIC spectra implies that no signal should be detectable at energies below 0.5 keV (wavelengths above 24.79 \AA) in the RGS data. We also note that emission lines on top of black-body-like emission have been detected in a number of X-ray sources with super-soft emission (either persistent or transient after a novae eruption). Those lines arise from resonant scattering of black-body emission by ambient electrons, mostly observable in high-inclination systems where the super-soft source is blocked \citep{2013A&A...559A..50N}. We can therefore still take optically thick, black body-like emission as the most likely interpretation of the emission of the softest X-rays.

\subsection{The SAS as a period of high $\dot{M}$}

\citet{2004A&A...415..609S} derived values for the accretion rate during a low state (from 1979 to 1989) of $\dot{M}_{low}=2\times10^{-9}$ \ms\ (d/914 pc)$^2$ yr$^{-1}$.  Throughout the SAS phase, an increase was noted in brightness across optical, UV, IR, and radio wavelengths \citep[][see references therein]{2019ApJ...884....8L}, while it diminished at higher energies. This pattern can only be understood if the accretion rate during the SAS rose compared to its quiescent state; any scenario where $\dot{M}$ stayed constant at the quiescent rate cannot account for the multiwavelength fluxes observed during the SAS. \citet{2018A&A...616A..53L} determined the accretion rate at the onset of the SAS, high state, to be $\dot{M}_{high} = 8.5 \times 10^{-8}$ \ms yr$^{-1}$ (d/914 pc)$^2$ based on \xmm/Obs1 data. The luminosity of the optically thick X-ray spectral component, indicative of the accretion rate, consistently maintained values around 10$^{35}$ \lumcgs during the SAS, as reflected by the \xmm\ spectral data (Fig. \ref{fig:xmm_evol}).

The multiwavelength data collected during the SAS provide strong evidence of its origin, which is related to a period of increased accretion in the accretion disk and its subsequent brightening. Those data are not compatible with a scenario in which the SAS would have been due to an episode of nuclear burning on the white dwarf surface, even if there was indeed super-soft X-ray emission \citep[in this state, \tcrb's X-ray spectrum is $\alpha/\delta$-type in the classification scheme proposed by][]{2013A&A...559A...6L}.\jls{} Its luminosity and temperature imply a size of the emitting region, R$_{bb}$, which is much smaller than the radius of a white dwarf, R$_{WD}$ (see Table \ref{tab:1}). The period of low flux at high energies (BAT, 15--50 keV) that started close to the onset of the SAS   cannot be attributed to enhanced nuclear burning on the WD surface either. It is, however, naturally explained by changes in the optical depth of the accretion disc's boundary layer during an episode of enhanced accretion rate.

The dense \swift/UVOT coverage also allows us to estimate the accretion rate during the SAS by measuring the Balmer jump intensity. During the SAS, we see that the Balmer jump went into emission (Fig. \ref{fig:uv_spec}). There can be several reasons for this, but the Balmer-jump emission is most likely dominated by H free-bound emission from continuum to level 2. 
%In addition, 
Scattering may cause additional emission below 3600$\AA$.\footnote{Work on this topic by \citet{1985MNRAS.212..231L} shows the effect of $b-f$ emission on the spectrum of the cataclysmic variable RU~Peg obtained with the IUE, where the spectrum is mostly flat over the Balmer jump. It is difficult to find good examples for larger Balmer jump emission, partly because IUE and HST UV observations do not cover far enough in the red. Continuum electron opacity effects have been discussed by \citet{1986ApJ...308..805K} in quasar spectra. Those can, in principle, also enhance the emission shortward of the Balmer jump. Significant Balmer continuum emission has been discussed in the AGN NGC~5548 \citep{2019MNRAS.489.5284K,2015A&A...575A..22M}, where a central source illuminates a region with column density $10^{23} cm^{-2}$, including all important continuum emission physics. Gas densities are typically $10^{10} cm^{-3}$. Despite the different scales used in their simple model, the radiative process is dependent on optical depth and local density, which makes this a credible approach to consider.}
We can estimate the flux in the Balmer continuum during the SAS. The flux on average is $1.5\times10^{-13}$ \fluxcgs higher over a region of 1300$\AA$\ wide giving a total flux of $1.95\times 10^{-10}$ \fluxcgs. Assuming a reddening of E(B-V) = 0.06 mag and using the reddening curve of \citet{cardelli_clayton_mathis} with R$_V$ = 3.1, we derived a correction factor of 0.36 and obtain a source luminosity in the Balmer continuum of $\approx7\times10^{33}\, {\rm erg\, s}^{-1}$, which is a factor of a few smaller than the unabsorbed X-ray luminosity. However, given the simplicity of the above calculation, we considered the Balmer jump to also support the nature of the SAS as due to an increase in the accretion rate.

\subsection{About the predictions for the next nova outburst.}

Two key white dwarf parameters are useful in determining the required ignition mass for nova eruptions:\ its mass and core temperature. These parameters do not change appreciably over a few accretion-nova cycles of a recurrent nova. Thus, for each recurrent nova system, we can consider the ignition mass to be a constant. Under that approximation, the recurrence intervals are determined as ignition mass divided by the average accretion rate.

It is therefore tempting to assume that the accretion rate in a periodic recurrent nova is roughly constant. However, everything we know about cataclysmic variables and symbiotic binaries shows that the accretion rate is never truly constant, the SAS of \tcrb\ being exhibit \#1. Each binary system probably has a typical accretion rate (hence the typical recurrence period) but with fluctuations around the mean. 

Among the Galactic recurrent novae, U~Sco has had the most known eruptions with eleven \citep[in 1863, 1906, 1917, 1936, 1945, 1969, 1979, 1987, 1999, 2010, and 2022;][and recent updates]{2010ApJS..187..275S}, for eruption intervals of 43, 11, 19, 9, 24, 10, 8, 12, 11, and 12 years.  If there are three missing eruptions between 1863 and 1906, and one between 1945 and 1969 (plausible but unproven), then U~Sco may be quite regular with approximately 10 year recurrence intervals. Even so, the average and standard deviation (of the interval values around 10) are 10.4 and 1.5 years, respectively.  Behind U~Sco, there are now nine known eruptions of RS~Oph (in 1898, 1907, 1933, 1945, 1958, 1967, 1985, 2006, and 2021) for eruption intervals of 9, 26, 12, 13, 9, 18, 21, and 15 years, with an average of 15.5 years and a standard deviation of 6.2 years, far less regular than U~Sco \citep[see also the case of M31N 2018-12a;][]{2018ApJ...857...68H}

Such scatter and irregularities must be taken into consideration when using the 80 year interval to predict the next eruption of \tcrb. If \tcrb\ is as irregular as U~Sco (in terms of the standard deviation and mean), the 1$\sigma$-uncertainty is about 10 years (and even larger if \tcrb\ is as irregular as RS~Oph). Similarly, while we believe states such as the SAS are necessary for the white dwarf to accumulate sufficient fuel for a nova eruption, it is premature to assume that the onset of SAS would be followed, exactly 8 years later, by a nova eruption. Now that \tcrb\ has gone through an SAS, the white dwarf might be close to having sufficient fresh fuel for a new nova eruption; however, the existing data do not allow a reliable prediction of its timing.

\section{Summary and conclusions} \label{sec:conc}

In this work, we  present a comprehensive multiwavelength analysis of the recurrent nova \tcrb, focusing on its recent SAS, which began in late 2014, followed by the deep optical/UV minimum observed in 2023. Our analysis is based on extensive observational campaigns with \xmm, \nustar, \swift, and \hst. Our principal findings can be summarized as follows:
\begin{itemize}
    \item The X-ray spectral evolution of \tcrb\ is characterized by the presence of a soft, optically thick emission component (modeled as a blackbody with kT $\approx$ 40 eV) during the SAS, which coexists with a multitemperature, optically thin plasma from the cooling flow. This soft component was not present during the 2025 recovery state.
    \item In our observations, we detected a steady decline (significant at a 2.58$\sigma$ level) in the luminosity of the optically thick emission, persisting until the conclusion of the SAS. This pattern aligns with a progressive decrease in $\dot{M}$ during this time frame, resulting in a boundary layer that gradually becomes less optically thick. 
    \item A detailed timing analysis of the \xmm/$pn$+MOS1+MOS2 and OM fast-mode light curves from multiple epochs (Obs1, Obs2, Obs3, Obs4, Obs5, and Obs6) reveals no significant, coherent periodicities. Specifically, we find no evidence for the $\sim$6000 s periodicity previously reported by \citet{zhekov19}.
\end{itemize} 
Based on these results, we draw the following conclusions:
\begin{itemize}
    \item We conclude that the SAS is driven by an increased mass accretion rate. This increased accretion rate makes the boundary layer optically thick, giving rise to the bright, soft X-ray blackbody component and partially suppressing the hard X-ray emission from the inner accretion flow. The small size of the blackbody emitting region from modeling of the supersoft X-ray component rules out the picture in which nuclear burning on the surface of the WD was the driver of the SAS.
    \item For the origin of the softest X-ray spectral component during the SAS, our inability to confirm the RGS X-ray spectral lines, initially claimed by \citet{2024MNRAS.532.1421T}, leads us to reaffirm that the supersoft X-ray spectral component is more consistent with optically thick blackbody emission than optically thin plasma emission.
    \item In line with this interpretation, we noted that the brightness of the optically thin component has consistently risen until the most recent observation taken with \xmm\ in July 2025, as reported in this study. This upward trend implies a lower accretion rate and the boundary layer's plasma predominantly becoming optically thin. The re-emergence of the hard X-ray component, witnessed by \swift/BAT during the faint state provides powerful confirmation of the model where the boundary layer transitions between optically thick and thin states. 
    \item The 2023 faint state is an intrinsic phenomenon that is unlikely to be related to an obscuration event caused by dust. 

\end{itemize}

While \tcrb\ clearly appears to be in the final stages before its next nova eruption, our analysis of its historical recurrence pattern suggests a significant scatter of at least 10 years around the mean $\sim$80-year period. Therefore, while an eruption is anticipated, the existing data do not allow for a precise prediction of its timing.

\begin{acknowledgements}
We acknowledge the anonymous referee for their careful reading and comments that significantly improved the manuscript. The authors acknowledge fruitful discussions with Eric Feigelson, Ole K\"onig, Gloria Sala, Steve Shore, Bradley Schaefer and Raymundo Baptista about different aspects of the manuscript. GJML is member of the CIC-CONICET (Argentina). JLS was supported by NASA grants 80NSSC25K7068 and 80NSSC25K7082. NPMK is funded by the UKSA. This work is based on observations obtained with \xmm\ an ESA science mission with instruments and contributions directly funded by ESA Member States and NASA. 
This research is based on observations made with the NASA/ESA Hubble Space Telescope obtained from the Space Telescope Science Institute, which is operated by the Association of Universities for Research in Astronomy, Inc., under NASA contract NAS 5–26555. These observations are associated with program \#16486.  This research has made use of data obtained with the \nustar\ mission, a project led by the California Institute of Technology (Caltech), managed by the Jet Propulsion Laboratory (JPL) and funded by NASA. This research has made use of the \nustar\ Data Analysis Software (NuSTARDAS) jointly developed by the ASI Science Data Center (ASDC, Italy) and the California Institute of Technology (Caltech, USA). We acknowledge use of the NASA HEASARC Swift tools as well as Astropy. We used of Physical Reference Data of the National Institute of Standards and Technology (NIST, USA). We thank the entire \swift\ team for accepting and planning our multiple Target-of-Opportunity requests. We thank Norbert Schartel for approving the \xmm\ DDT request. We acknowledge with thanks the variable star observations from the AAVSO International Database contributed by observers worldwide and used in this research. 
\end{acknowledgements}

\bibliographystyle{aa}
\bibliography{listaref_MASTER}

\begin{appendix} 

\section{Observation logs and X-ray spectra modeling results \label{appen:sec1}}

The resulting parameters of the X-ray spectral modeling are presented in Table \ref{tab:1}.

\begin{sidewaystable}[!htbp]
\centering
\begin{minipage}{\textheight}
\caption{Parameters from fitting the \xmm, \nustar\ and \swift\ spectra with a model \texttt{TBabs$\times$(partcov$\times$TBabs)(bbodyrad + mkcflow + gaussian)}.
Unabsorbed luminosities are calculated in the 0.001-50 keV energy band. Luminosity and $\dot{M}$ are determined assuming a distance of 914 pc. Elemental abundances are quoted in units of abundances from \citet{2000ApJ...542..914W}. Statistical errors are calculated at the 90\% confidence level. \label{tab:1}}
\renewcommand{\arraystretch}{1.}
\begin{tabular}{lccccccccc}
\hline\hline
& Obs1 & Obs2 & Obs3+Nu1 & Obs4 & Obs5 & Obs6 & Nu2+$Swift$ & Nu3+$Swift$ & Obs7 \\ 
\hline
Soft Count rate \tablefootmark{1} & 0.108& 0.040 & 0.0064 &  0.0032 & 0.027 & 0.0088 &  0.003 & 0.006 & 0.002 \\
Hard Count rate\tablefootmark{2} & 0.039 & 0.076 & 0.155 & 0.241 & 0.148 & 0.401 & 0.67/0.015 & 0.21/0.018 & 0.35\\
N$_{H}$ (F.C.) [10$^{22}$ cm$^{-2}$] \tablefootmark{3}&0.05$\pm$0.01 & 0.14$\pm$0.07 & 0.7$_{-0.4}^{+0.7}$ & 0.07$\pm$0.03 & 0.12$\pm$0.01 & 0.11$\pm$0.01  & 19$\pm$10 & 43$\pm$15 & 0.07$_{-0.06}^{+0.09}$\\ 
N$_{H}$ (P.C.) [10$^{22}$ cm$^{-2}$] \tablefootmark{4} & 68$\pm$3 & 46$\pm$3 & 29$\pm$2 & 53$\pm$3 & 29$\pm$1 & 40$\pm$1 & 65$_{-16}^{+40}$ & 157$_{-45}^{+83}$ & 45$\pm$1 \\
Covering fraction & 0.997$\pm$0.001 & 0.993$\pm$0.002 & 0.994$_{-0.004}^{+0.002}$ & 0.999$\pm$0.001 & 0.993$\pm$0.002 & 0.998$\pm$0.001 & 0.6$\pm$0.2 & 0.7$\pm$0.1 & 0.999 (fixed)\\
kT$_{bb}$[keV] & 0.034$\pm$0.002 & 0.037$\pm$0.005 & 0.04$\pm$0.02 & 0.041$\pm$0.015 & 0.029$\pm$0.001 & 0.04$\pm$0.01 & \nodata & \nodata & \nodata  \\
R$_{bb}$[km]  & 2084$\pm$500 & 1000$\pm$20 & 1200$\pm$200 & $<$ 500 & 3200$\pm$200 & 450$\pm$200  & \nodata & \nodata & \nodata \\ 
kT$_{\rm max}$ (keV) & 12$\pm$3 & 28$_{-2}^{+3}$ & 24$^{+8}_{-5}$ & 33$\pm$3 & 52$\pm$5 & 48$\pm$4  & 54$_{-5}^{+7}$ & 35$_{-8}^{+10}$ & 40$\pm$5 \\
Z/Z$_{\odot}$ & 1.5$\pm$0.5 & 3.1$\pm$0.8 & 1.9$\pm$0.5 & 1.7$\pm$0.2 & 4$\pm$0.4 & 2.8$\pm$0.5 & 1.0$\pm$0.2 & 1.0$\pm$0.4 & 1.0$\pm$0.1\\
$\dot{M}_{thin}$ [10$^{-9}$ M$_{\odot}$ yr$^{-1}$] & 0.22$\pm$0.07 & 0.11$\pm$0.05 & 0.21$\pm$0.06 & 0.4$\pm$0.1 & 0.17$\pm$0.03 & 0.38$\pm$0.02 & 1.2$\pm$0.1 & 1.2$\pm$0.5 & 0.42$\pm$0.01 \\
$\chi^{2}$/d.o.f & 210/226 & 87/84 & 211/196 & 487/455 & 280/197 & 484/464 & 574/561 & 203/211 & 779/712 \\
Log [L$_{bb}$]  & 35.9$\pm$0.7 & 35.5$\pm$0.7 & 35.9$\pm$0.8 &  35.4$\pm$1.2 & 36$\pm$1  & 34.8$\pm$1.1 & \nodata & \nodata & \nodata   \\
Log [L$_{cf}$]  & 32.74$\pm$0.03 & 32.92$\pm$0.03 & 33.12$\sim$0.02 & 33.50$\pm$0.03 & 33.12$\pm$0.03 & 33.65$\pm$0.03 & 34.20$\pm$0.02 & 34.02$\pm$0.02 & 33.61$\pm$0.02 \\
\hline
\end{tabular} 

\tablefoot{
\tablefoottext{1}{Count rate in the 0.3-1.0 keV energy range. In the case of \xmm\ we list the $pn$ count rate. In this energy range we also list the \swift/XRT count rate in the case of \nustar+\swift\ spectra.}
\tablefoottext{2}{$pn$ camera count rate in the 1-10 keV energy range in the case of \xmm. \swift\ count rate in the 1-10 keV energy range while \nustar\ count rate were computed in the 3-50 keV energy range.}
\tablefoottext{3}{Absorption column of the full covering absorber model, \texttt{TBabs}.}
\tablefoottext{4}{Absorption column of the partial covering absorber model, \texttt{partcov$\times$TBabs}.}
}
\end{minipage}
\end{sidewaystable}

\section{Auto-regressive analysis of X-ray and OM light curves \label{appen}}

This appendix presents the detailed results of the time-series analysis performed on the \xmm\ X-ray (EPIC) and optical (OM) light curves to search for coherent periodicities. As described in Sect. \ref{sec:timing}, our method involves a two-stage process to robustly distinguish periodic signals from the strong, aperiodic red noise characteristic of accreting systems.\\
The following figures (Fig. \ref{fig:om_2018} through \ref{fig:ar_2022b}) show the results of this procedure for each analyzed observation. The optical brightness during Obs1 and Obs2 led to strong saturation effects in the OM fast mode data, and the OM was blocked during Obs7 for the same reason; therefore, we only present the timing analysis for the OM data from Obs3, Obs4, Obs5, and Obs6. For the X-ray data, we analyzed all observations from Obs2 to Obs6.\\
Each figure consists of seven panels that provide a comprehensive diagnostic of the analysis:
(a) The raw, barycenter-corrected light curve; (b) \& (c) The Autocorrelation and Partial Autocorrelation Functions of the raw data, typically showing the exponential decay characteristic of red noise; (d) \& (e) The ACF and PACF of the "whitened" residuals after subtracting the best-fit autoregressive (AR) model. The lack of significant correlation demonstrates the successful removal of the red noise; (f) The power spectral density (PSD) of the original data (black) overlaid with the PSD of the fitted AR model (red), showing the model's accuracy in capturing the red noise shape; (g) LS power spectrum of the whitened residuals. The horizontal dashed line indicates the 99.99\% False Alarm Probability (FAP) level. No peaks are observed above this threshold in any observation, indicating the absence of statistically significant periodicities. The location of the previously reported $\sim$6000-second periodicity is unmarked as no power is present at or near this frequency. 

\begin{figure*}
\begin{center}
\includegraphics[scale=0.6]{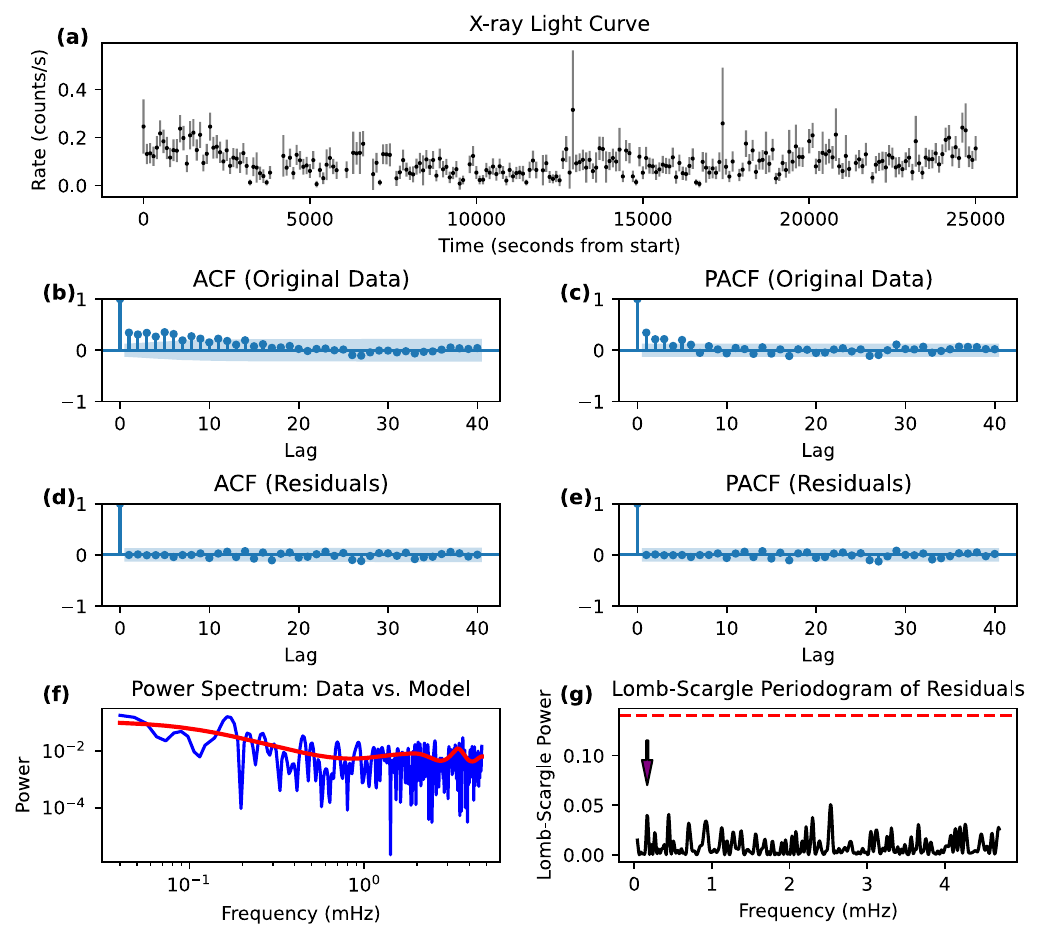}
\caption{Autoregressive processing of the Obs2 \xmm/$pn$+MOS1+MOS2 observation.} 
\label{fig:om_2018}
\end{center}
\end{figure*}

\begin{figure*}
\begin{center}
\includegraphics[scale=0.6]{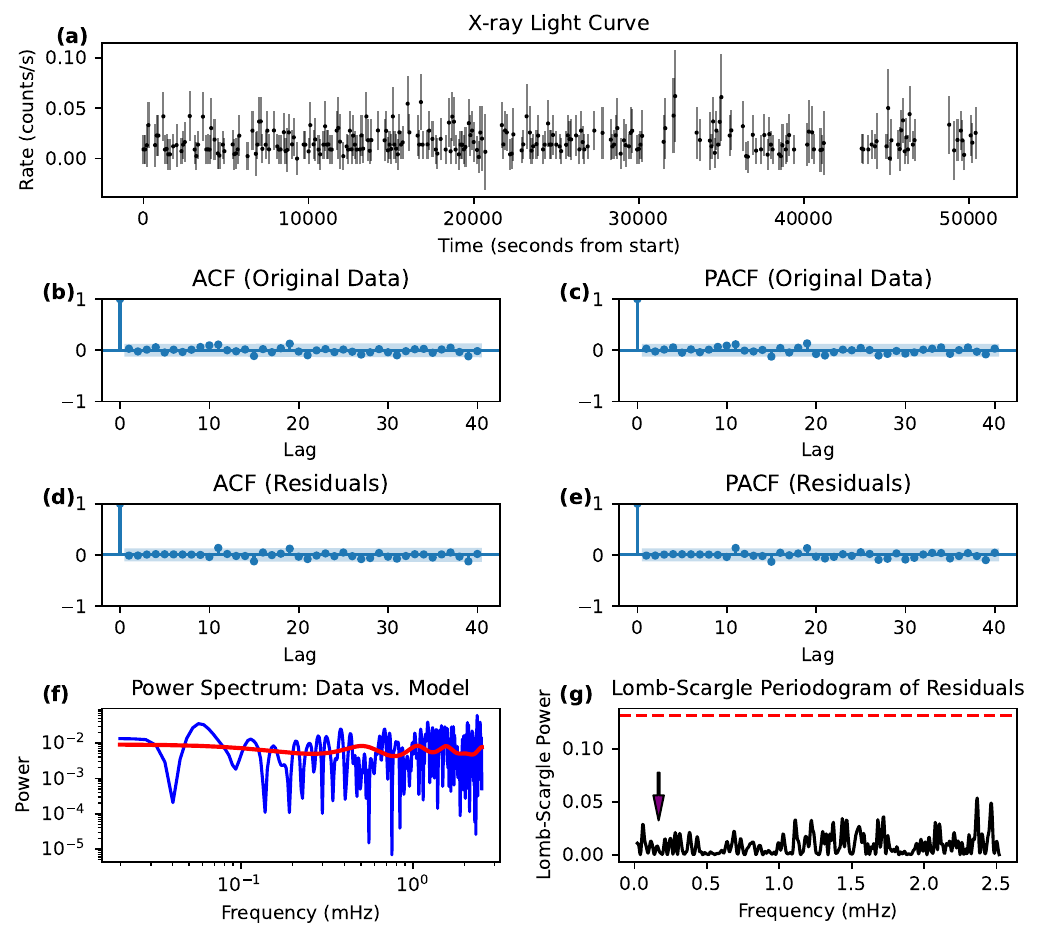}
\caption{Autoregressive processing of the Obs3 \xmm/$pn$+MOS1+MOS2 observation. }
\label{fig:om_2020}
\end{center}
\end{figure*}

\begin{figure*}
\begin{center}
\includegraphics[scale=0.6]{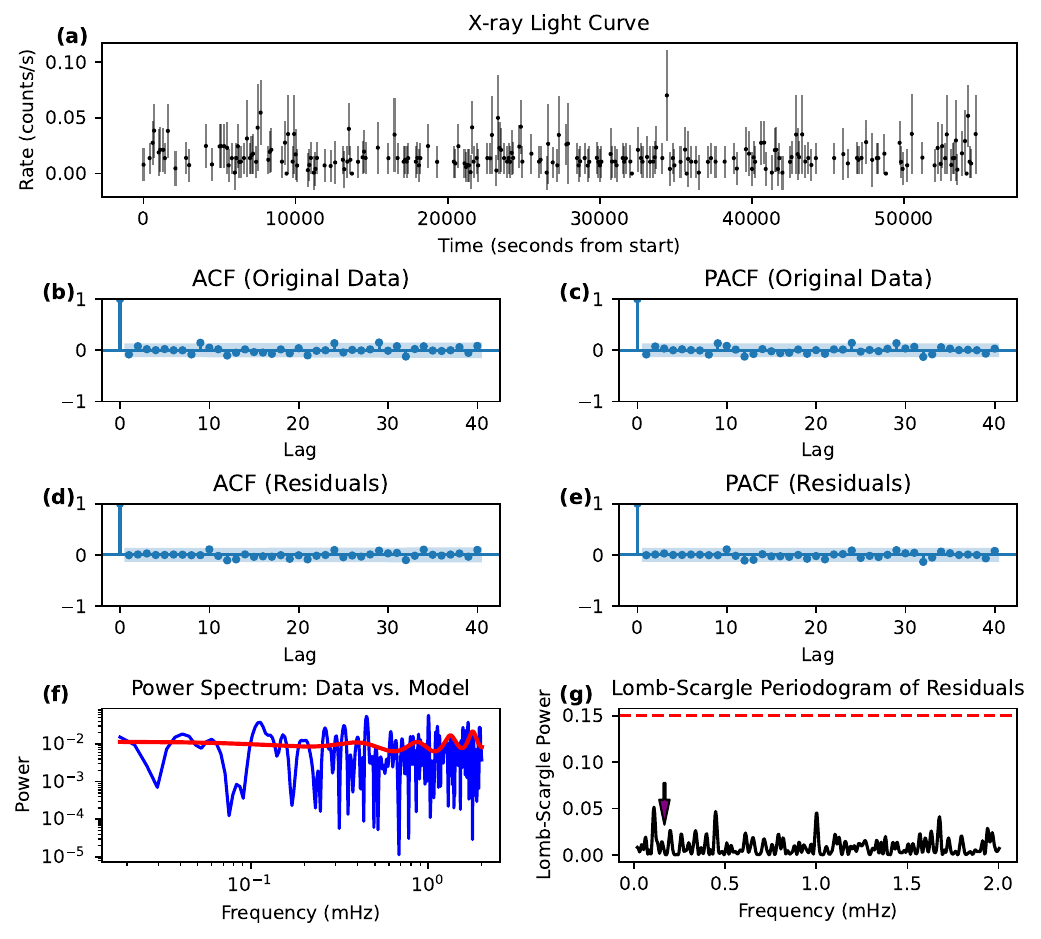}
\caption{Autoregressive processing of the Obs4 \xmm/$pn$+MOS1+MOS2 observation.}
\label{fig:om_2021}
\end{center}
\end{figure*}

\begin{figure*}
\begin{center}
\includegraphics[scale=0.6]{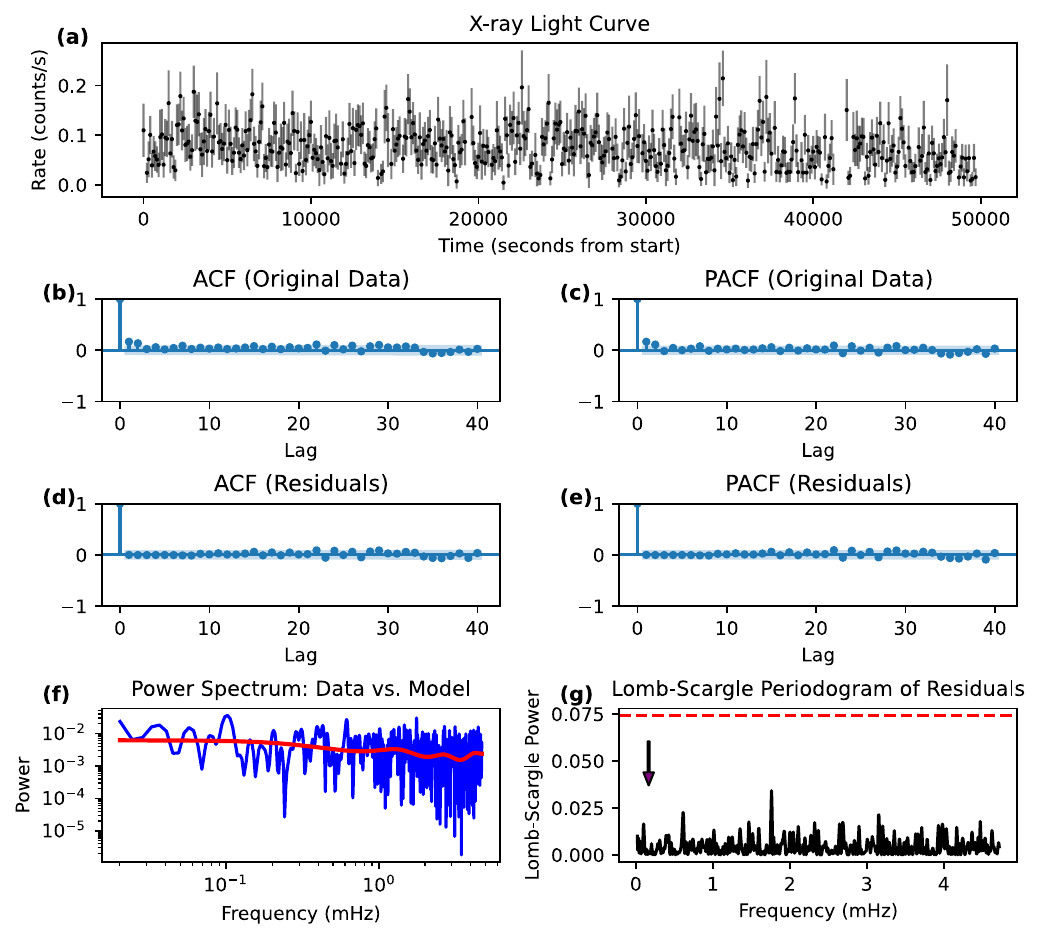}
\caption{Autoregressive processing of the Obs5 \xmm/$pn$+MOS1+MOS2 observation. }
\label{fig:om_2022a}
\end{center}
\end{figure*}

\begin{figure*}
\begin{center}
\includegraphics[scale=0.6]{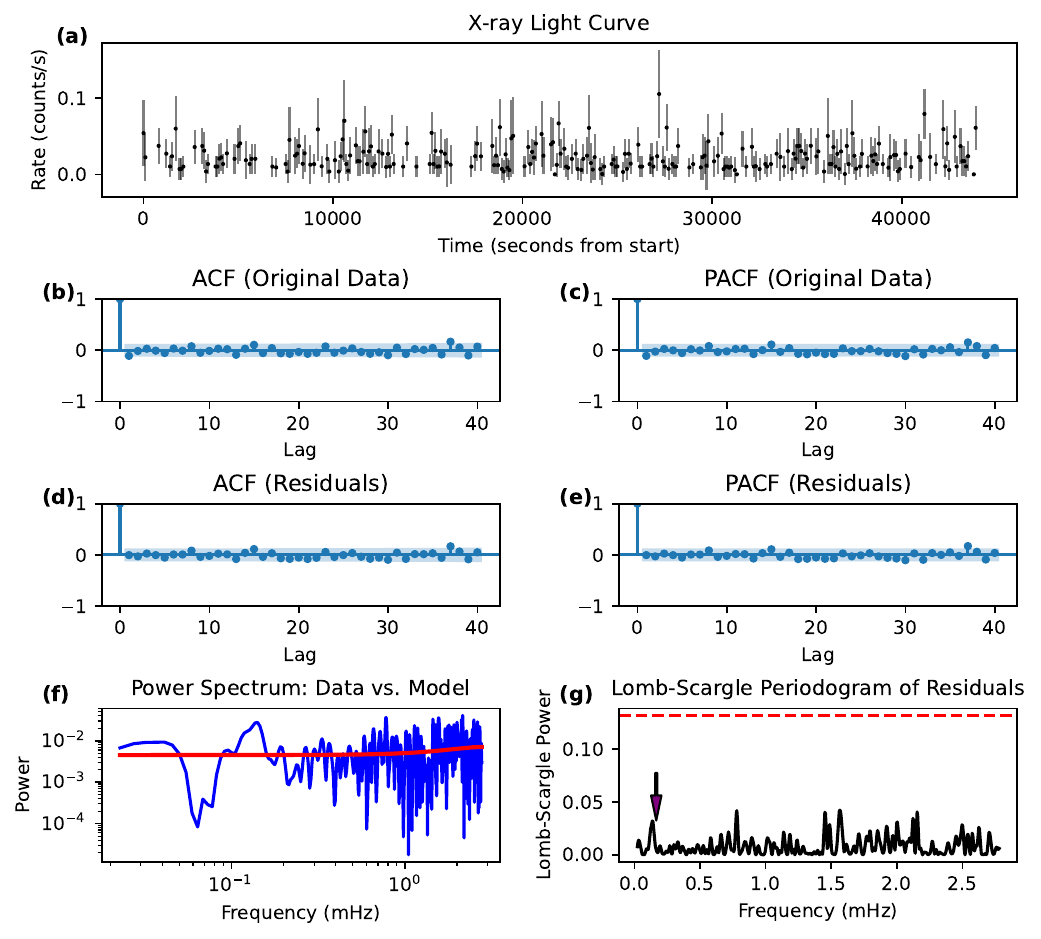}
\caption{Autoregressive processing of the Obs6 \xmm/$pn$+MOS1+MOS2 observation. }
\label{fig:om_2022b}
\end{center}
\end{figure*}

%-------------------------------------------------------- OM
\begin{figure*}
\begin{center}
\includegraphics[scale=0.6]{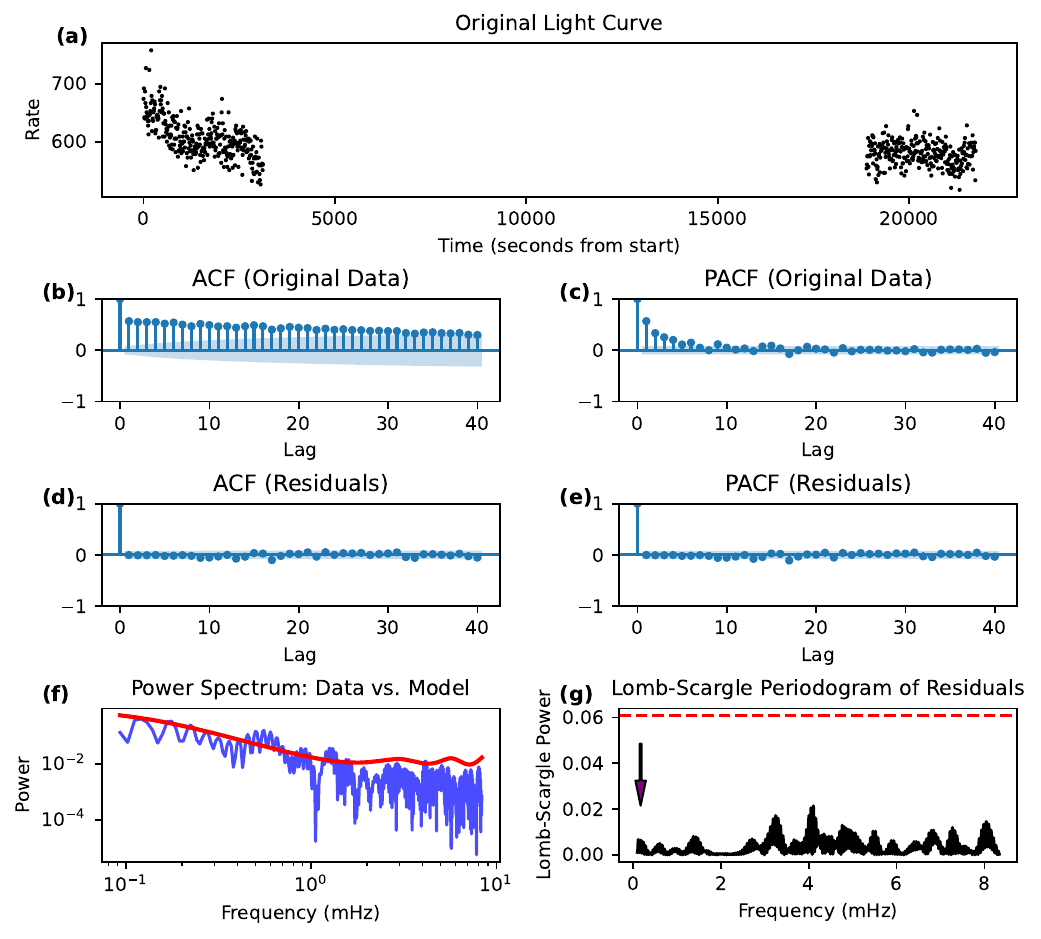}
\caption{Autoregressive processing of the Obs1 \xmm/OM observation. } 
\label{fig:ar_2017}
\end{center}
\end{figure*}

\begin{figure*}
\begin{center}
\includegraphics[scale=0.6]{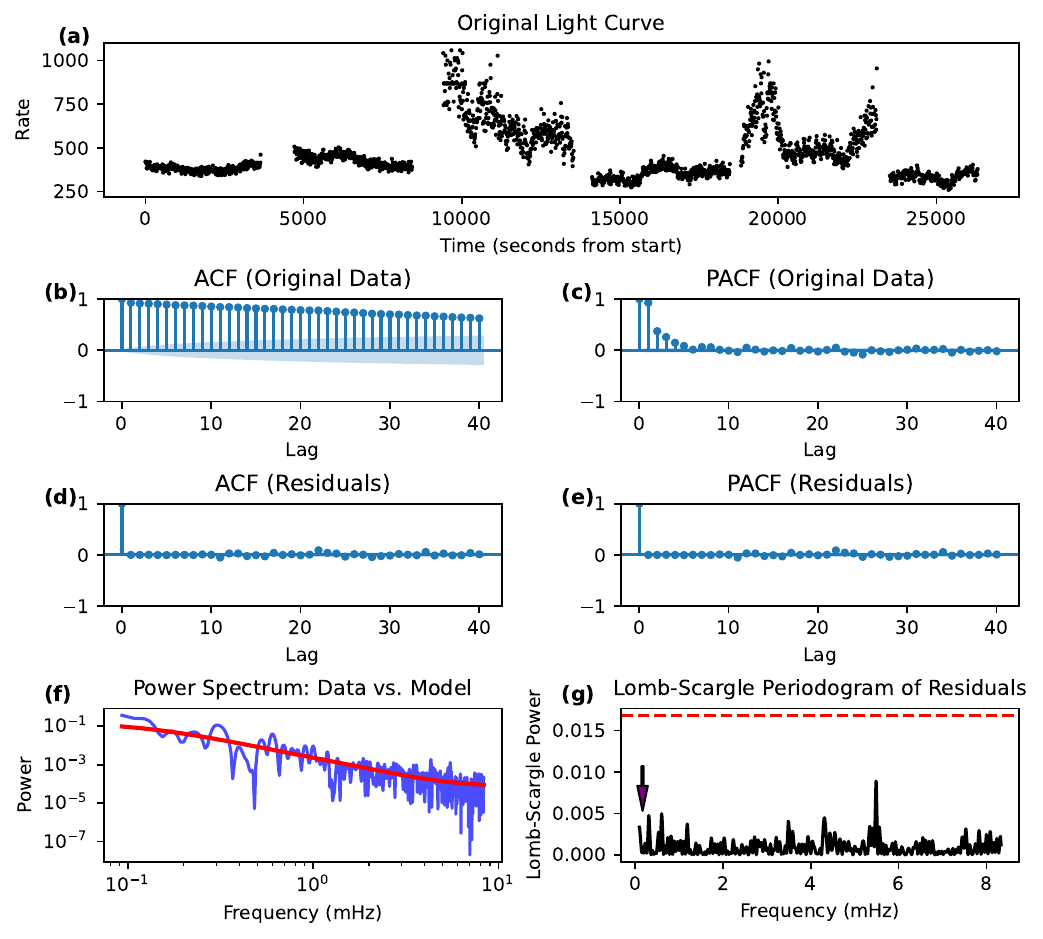}
\caption{Autoregressive processing of the Obs2 \xmm/OM observation. } 
\label{fig:ar_2018}
\end{center}
\end{figure*}

\begin{figure*}
\begin{center}
\includegraphics[scale=0.6]{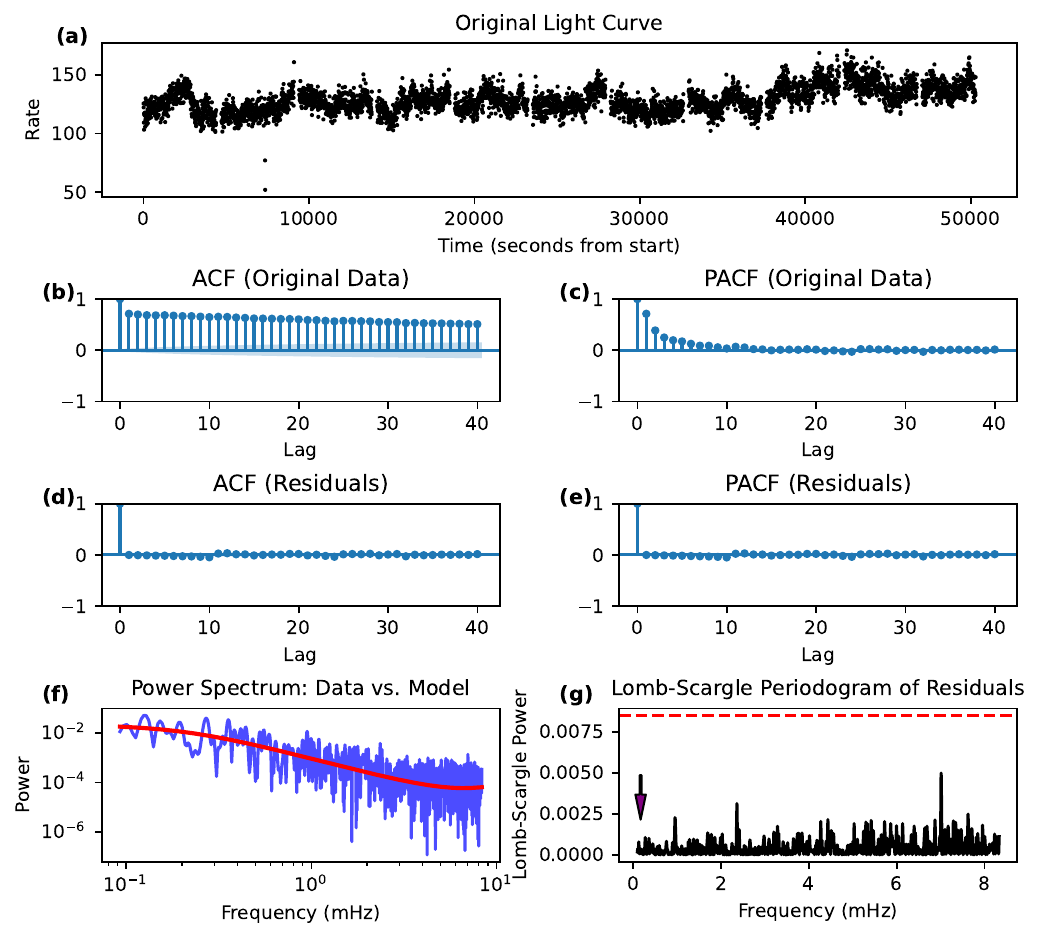}
\caption{Autoregressive processing of the Obs3 \xmm/OM observation. }
\label{fig:ar_2020}
\end{center}
\end{figure*}

\begin{figure*}
\begin{center}
\includegraphics[scale=0.6]{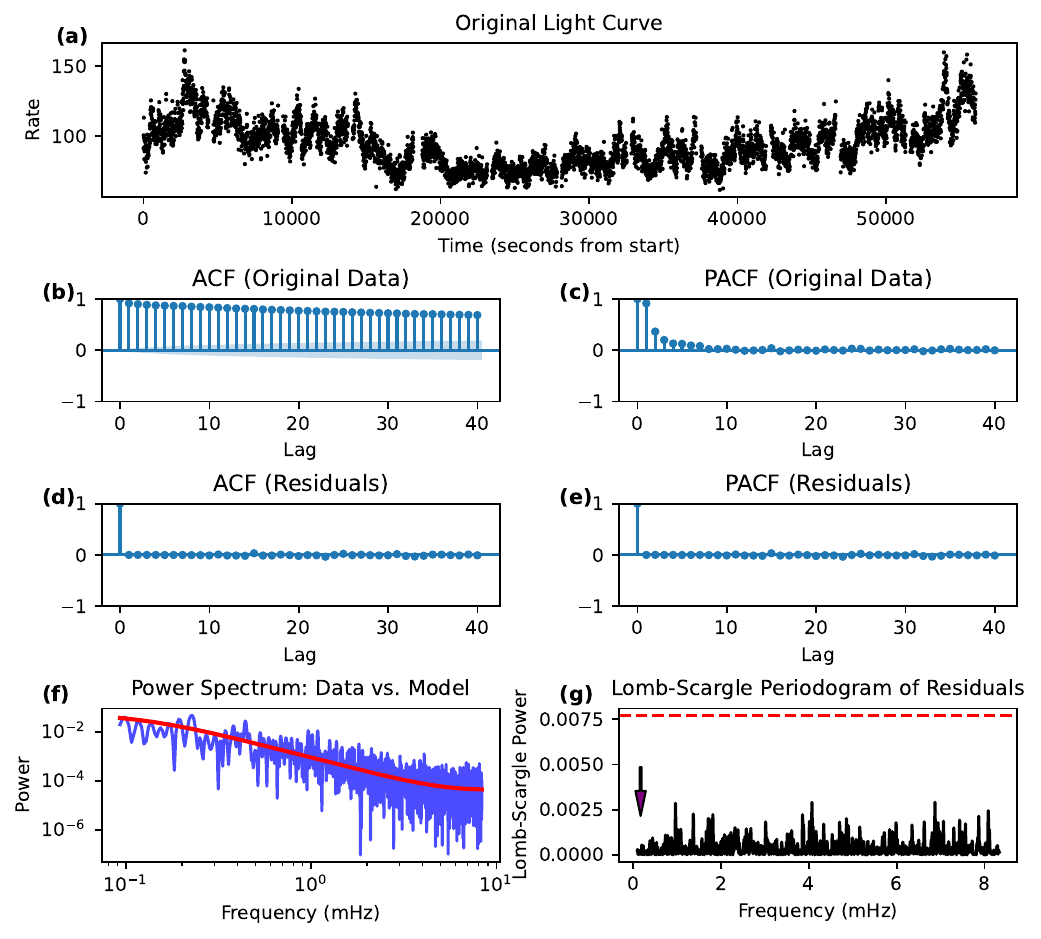}
\caption{Autoregressive processing of the Obs4 \xmm/OM observation.}
\label{fig:ar_2021}
\end{center}
\end{figure*}

\begin{figure*}
\begin{center}
\includegraphics[scale=0.6]{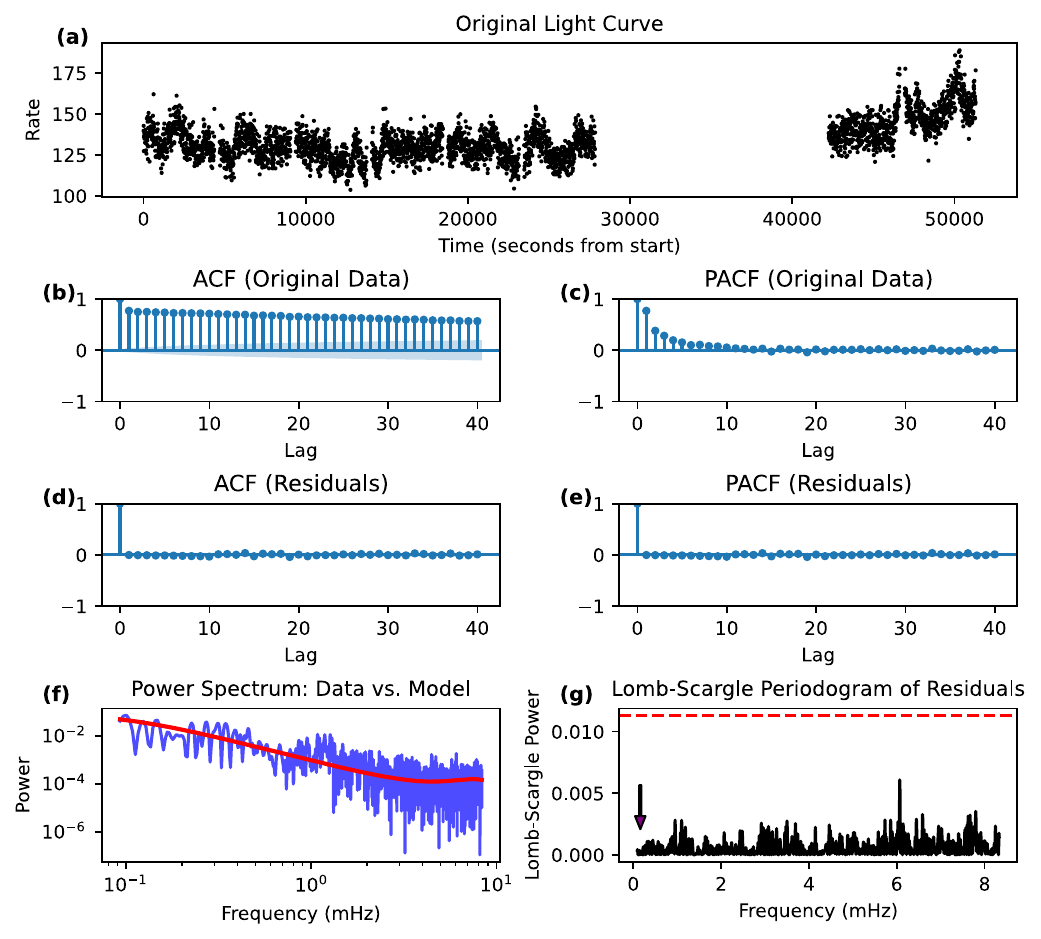}
\caption{Autoregressive processing of the Obs5 \xmm/OM observation. } 
\label{fig:ar_2022a}
\end{center}
\end{figure*}

\begin{figure*}
\begin{center}
\includegraphics[scale=0.6]{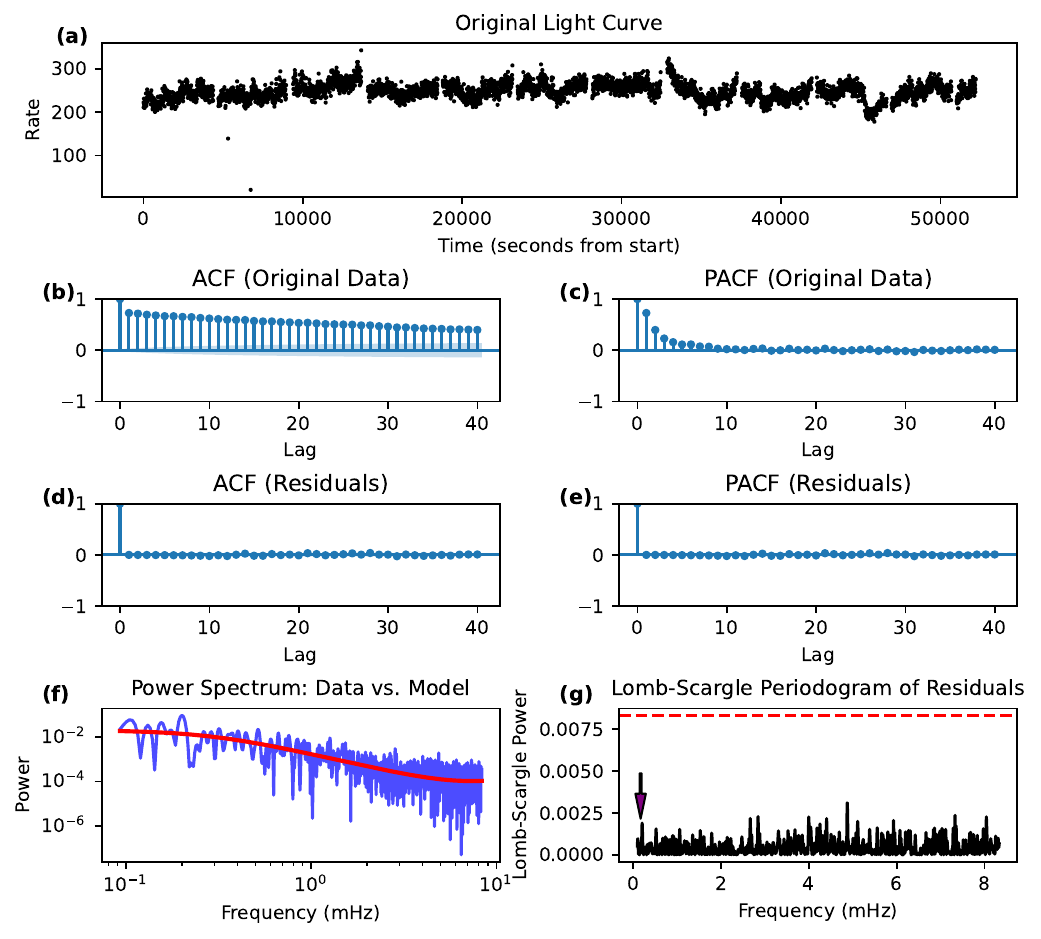}
\caption{Autoregressive processing of the Obs6 \xmm/OM observation. }
\label{fig:ar_2022b}
\end{center}
\end{figure*}

\end{appendix}

\end{document}